\documentclass[iop]{emulateapj} 

\usepackage{color}

\def\h2o  {H$_2$O}
\def\ho {H$_0$}
\def\WISE{\emph{WISE}}

\begin{document} 
 
 \title{Enhancing the H$_{2}$O Megamaser Detection Rate Using Optical and Mid-infrared Photometry}

 \author{C. Y. Kuo\altaffilmark{1}, A. Constantin\altaffilmark{2}, J. A. Braatz\altaffilmark{3},
Chung, H. H.\altaffilmark{1}, C. A. Witherspoon\altaffilmark{2}, D. Pesce\altaffilmark{4}, C. M. V. Impellizzeri\altaffilmark{3,5}, F. Gao\altaffilmark{6},
Lei Hao\altaffilmark{7}, J.-H. Woo\altaffilmark{8}, Ingyin Zaw\altaffilmark{9}}
 
\affil{\altaffilmark{1} Physics Department, National Sun Yat-Sen University, No. 70, Lien-Hai Rd, Kaosiung City 80424, Taiwan, R.O.C }    
\affil{\altaffilmark{2} Department of Physics and Astronomy, James Madison University, Harrisonburg, VA 22807, USA }                                            
\affil{\altaffilmark{3} National Radio Astronomy Observatory, 520 Edgemont Road, Charlottesville, VA 22903, USA}                                                    
\affil{\altaffilmark{4} Department of Astronomy, University of Virginia, Charlottesville, VA 22904}
\affil{\altaffilmark{5} Joint Alma Office, Alsonso de Cordova 3107, Vitacura, Santiago, Chile}         
\affil{\altaffilmark{6} Max Planck Institute for extraterrestrial Physics, Giessenbach str., 85748, Garching, Germany}
\affil{\altaffilmark{7}Shanghai Astronomical Observatory, Chinese Academy of Sciences Shanghai 200030, P.R. China}
\affil{\altaffilmark{8} Astronomy Program, Department of Physics and Astronomy, Seoul National University, Seoul 151-742, Korea  }                                   
\affil{\altaffilmark{9} New York University Abu Dhabi, Abu Dhabi, UAE }

\begin{abstract}
Water megamasers from circumnuclear disks in  galaxy centers provide the most accurate measurements of supermassive black hole masses and  uniquely probe the sub-parsec accretion processes.  At the same time, these systems offer independent crucial constraints of the Hubble Constant in the nearby universe, and thus, the arguably best single constraint on the nature of dark energy.   The chances of finding these golden standards are however abysmally low, at an overall $\la 3$\% for any level of water maser emission detected at 22 GHz, and $\la 1$\% for those exhibiting the disk-like configuration.   We provide here a thorough summary of the current state of the detection of water megamaser disks, along with a novel investigation of the likelihood of increasing their detection rates based on a multivariate parameter analysis of optical and mid-infrared photometric properties of the largest database of galaxies surveyed for the 22 GHz emission.   
We find that galaxies with water megamaser emission tend to associate with strong emission in all \emph{WISE} mid-infrared wavelengths, with the strongest enhancement in the $W4$ band, at 22$\micron$, as well as with previously proposed and newly found indicators of AGN strength in the mid-infrared, such as red $W1-W2$ and $W1-W4$ colors, and the integrated mid-infrared luminosity of the host galaxy.  These trends offer a potential boost of the megamaser detection rates to $6 - 15$\%, or a factor of $2-8$ relative to the current rates, depending on the chosen sample selection criteria, while fostering real chances for discovering $\ga 20$ new megamaser disks. 


\end{abstract} 
 

 
\section{Introduction} \label{intro}

Astrophysical water masers are natural \textit{microwave amplifiers by stimulated emission of radiation}. When detected in galaxy centers and are extremely luminous, i.e. millions of times more luminous (brightness temperatures to $10^{16}$ K; isotropic luminosities $L_{\rm H_{2}O} \ga 10 L_{\sun}$) than those associated with typical star-forming regions in the spiral arms of our own Milky Way, they are termed ``megamasers."   
For a fraction ($\sim$ 30\%) of them,  disk-maser candidates are identified via the high velocity features (redshifted and blueshifted features) that their 22 GHz spectra show in addition to the systemic velocity components  (e.g., Kondratko et al. 2006, Kuo et al. 2011). 
Water megamasers in a disk-like configuration are a  treasure trove, for they advance our understanding of two extremely important issues: i) the current cosmological model and the nature of Dark Energy (DE), and ii) the galaxy formation and evolution processes.    

Vital constraints on the current cosmological model need an extremely accurate Hubble Constant \ho ~value.
The exquisite \emph{WMAP} cosmic microwave background radiation and \emph{Planck} data from redshifts $z \approx 1100$, when DE was negligible, are consistent with a wide range of values of \ho\  if one does not assume that the universe is flat, i.e., the current ``concordance" model (Spergel et al. 2007; Adam et al.  2016).    There is however tension between predicted values of \ho\ from \emph{Planck} and observed values based on standard candles (e.g., Riess et al. 2016) or strong lensing (Bovin et al. 2017).   Thus, independent methods of measuring \ho\ are needed in order to minimize systematic uncertainties, and ultimately to provide strong constraints of the geometry and the fraction ($\Omega_m$) of the critical density contributed by matter.  A 3\% or better accuracy in measuring \ho\ would arguably provide the best single constraint on the nature of DE where its effect is greatest (Hu 2005; Olling 2007).   One way of reaching this level of accuracy is, for example, to measure $\sim$10\% distances to each of  $\approx 10$ galaxies at distances $\sim 50 - 200$ Mpc (e.g., Greenhill 2004).      
To accomplish the goal we must, however, first identify these precious maser disks, and this requires a substantial increase in their current detection efficiency, which is one of the main objectives of this study.  


Regarding the galaxy evolution processes, the megamaser disks provide essential means of constraining the supermassive black hole (SMBH) occupation fraction, along with the universality (i.e., the overall shape and scatter) of the empirically found black hole -- host galaxy scaling relations\footnote{the $M_{\rm BH} -\sigma^*$ relation; Gebhardt et al. (2000); Ferrarese \& Merritt (2000); Tremaine et al. (2002); G\"{u}ltekin et al. (2009), Graham et al. (2011); and the  $M_{\rm BH} -L_{\rm bulge}$ relation, e.g., Marconi \& Hunt (2003) ~~~}, especially on the lower end of the distribution ($M_{\rm BH}  \leq 10^7 M_{\sun}$), where very few accurate  estimates are available (e.g., Greene et al. 2010; Greene et al. 2016).    Due to their well-measured Keplerian rotation, the megamaser disks are an optimal dynamical tracer of the central mass lurking in galaxy centers.  Accurate maser-based BH mass measurements via Very Long Baseline Interferometry (VLBI) observations  
do not suffer from biases from non-virial gas support (e.g., due to turbulence; Barth et al. 2001).  Moreover, unlike stellar dynamical methods, the VLBI measurements are not sensitive to assumptions about galaxy anisotropy and radial variation in dark matter fraction (which are not well constrained observationally; Gebhardt \& Thomas 2009; van den Bosch \& de Zeeuw 2010).   

Unfortunately, of $^>_{\sim} ~4000$ galaxy nuclei surveyed so far for water maser emission in 22 GHz, only $\sim160$ are detected, with $\sim30$\% of them possibly originating in disks.   The latest data on water maser surveys are curated and made publicly available by the Megamaser Cosmology Project (MCP)\footnote{https://safe.nrao.edu/wiki/bin/view/Main/MegamaserCosmologyProject}.  Only $\sim40$  of all these maser galaxies  show ``triple" spectra, i.e., displaying systemic and high-velocity emission in three line complexes, and only $\sim10$ are possibly good candidates for distance measurements, depending on the collecting area available for VLBI; only $<9$ of these objects lie at distances $D > 50$ Mpc, where their peculiar motions are a small fraction of their total motions, allowing thus estimates of \ho\   directly from maser distances and recessional velocities.  (Kuo et al. 2011, 2013, 2015, Reid et al. 2013, Gao et al. 2016).    


Good megamaser disk candidates are however extremely few.  The need to have the disk structure 
strong enough to enable VLBI imaging throughout the velocity ranges of each complex, combined with the requirement for not only the presence of very dense molecular gas, but also a suitable viewpoint, with the nuclear disk seen almost edge-on, cause the probability of detecting the ``right" megamaser to drop significantly.  
Detailed studies of these few precious maser disks clearly identify a parsec-scale, Keplerian rotation profile, and accelerations measured with the \emph{Green Bank Telescope (GBT)} and the VLBI mappings, together with refinements in the acceleration modeling and a 3-D disk fitting (Humphreys et al. 2008, 2013) already offer a $\sim$10\% uncertainty in \ho\ (e.g., Braatz et al. 2010, Kuo et al. 2011, Reid et al. 2012, 2013).  To increase the accuracy to $<$10\%, the current suite of telescopes needs an improvement in sensitivity by a factor of $\sim 2$, which will not be possible for at least another few decades.

\subsection{Megamasers and Their Hosts: Optical Studies and Their Limitations.} \label{limitations}

Initial surveys for maser activity in extragalactic sources have been closely driven by the idea that the maser emission must be associated with intense star-formation, and thus targeted objects with large infrared emission or unusually active, radio-loud, or morphologically peculiar galaxies.   Maser emission was detected, albeit extremely rarely, at the rate of $<1$ in $\sim 40$ targets (Henkel et al. 1984; Batchelor, Jauncey \& Whiteoak 1982; Whiteoak \& Gardner 1986; Henkel, Wouterloot \& Bally 1986; Hagiwara, Diamond \& Miyoshi 2002, 2003; Nakai, Sato \& Yamauchi 2002).   These searches revealed that the very few powerful megamasers found were hosted by galaxies with actively accreting SMBHs in their centers (i.e., Active Galactic Nuclei, AGNs), where the maser emission originates within about 1pc (3 light-years) or less of the nucleus (e.g., Claussen \& Lo 1986; Haschick et al. 1994); these findings led more recent searches for \h2o ~megamasers to focus on AGNs.  Improved sensitivity and newer spectrometers resulted in new detections in AGNs, however, the success rate remained a mere $\sim 3-5$\% (Braatz, Wilson \& Henkel 1996, 1997; Braatz et al. 2004; Greenhill et al. 2003; Kondratko et al. 2006; Braatz \& Gugliucci 2008; Braatz et al. 2010).   

To date, almost all megamaser disk sources were found in galaxies that only show narrow emission line activity in their centers (i.e., type 2 AGNs, which include both Seyferts and some Low Ionization Nuclear Emission Regions, or LINERs);  with extremely few exceptions, AGNs with broad emission (type 1) do not appear to host masers.  The standard interpretation for the difference between broad and narrow line emission AGNs is known as the ``Unification scenario" (Antonucci \& Miller 1985), where a dusty torus surrounds the nucleus and obscures the central engine (BH, accretion disk, and the broad line emission region) for large inclination angles.   Because the presence of water molecules requires a dusty environment, it has thus been inferred that masers might trace molecular material associated with this torus, or with an accretion disk that feeds the nucleus.  Consequently, the maser emission is expected to also be heavily geometry-dependent in the same manner as the type1/type 2 AGN classes, i.e., to occur for certain, very restricted, viewing angles, perhaps where the line of sight is along the plane of the molecular disk or torus.   This scenario appears to be confirmed in great detail by VLBI observations of the megamaser in NGC 4258 (Miyoshi et al. 1995; Greenhill et al. 1995; Lo 2005).   
Thus, there are good reasons for searching for megamasers in type 2 AGNs.

Nevertheless, {\sl there is not enough reason to restrict the megamaser search to narrow-lined AGNs, or to previously identified AGNs, for that matter.}   While there is increasing evidence that the maser disk emission is very likely to originate from a geometrically thin disk made of a large number of molecular clouds that connect the inner edge of the torus and the outer part of the accretion disk (e.g., Masini et al. 2016), there are only very few special cases\footnote{Hutchings et al. 1998; Crenshaw et al. 2000; Crenshaw \& Kraemer 2000; Ruiz et al. 2001; Das et al. 2005,2006; Fischer et al. 2010; Crenshaw et al. 2010; Fischer et al. 2011} for which information constraining the inclinations of individual AGN accretion disks relative to that of the surrounding torus is available.   
The fact that, e.g.,  the \h2o ~megamaser (albeit not a disk) emitting quasar MG J0414+0534 shows broad emission (Impellizzeri et al. 2008) proves that broad-lined  AGNs could also host megamasers.  Whether all powerful masers host a disk or not, remains still to be identified, however, the presence of a torus does not seem to play as crucial a role in amplifying the radiation as originally thought.  




Detailed characterization of the optical properties of the H$_{2}$O maser host galaxies (Zhu et al. 2011 and Constantin 2012) provide promising ways of identifying more concrete criteria of searching for \h2o ~megamasers.   These studies also show somewhat surprising trends in their host properties: megamasers are preferentially found in galaxies with strong intrinsic [O III] $\lambda$5007 emission and high levels of obscuration as measured by Balmer decrements,  while they tend to favor a possible ``goldilocks" range in other properties, e.g.: 1) a narrow range of $L_{[O III]}/{\sigma^*}^4$ values, or Eddington ratios, which suggests a narrow range in their BH accretion rates (concentrated at $L/L_{\rm Edd} \sim 10^{-2}$), and 2) a narrow distribution of [S II] $\lambda$6716/ $\lambda$6731 line ratios which implies narrow ranges in the electron densities in the line-emitting gas.   
Maser searches tailored on these findings can potentially result in a $\sim4$-fold increase of the maser detection rate (to $\ga 11$\%, from the current $<3$\%).   Unfortunately, there are very few spectroscopically confirmed galaxies inside this box of properties (e.g., in the Sloan Digital Sky Survey; SDSS), and this sample has been somewhat exhausted (e.g., discussion in van den Bosch et al. 2016). 

Interestingly, it is also possible that the high maser detection rate among Seyferts  ($\sim 8$\%) may not be due entirely to a genuine connection to this type of activity, as it is enhanced by biases associated with the recent maser surveys that targeted almost exclusively type 2 Seyferts (Constantin 2012).  Also, the overall maser detection rate is non-zero for non-AGN types; in fact, there is an equal probability to find masers among Transition Objects (that straddle the borders between starburst galaxies and AGNs in emission-line diagnostic diagrams; e.g., Baldwin et al. 1981; Veilleux \& Osterbrock 1987; Kewley et al. 2006) as in LINERs ($\sim$1.6\%), and it is twice as likely to find them in Star-forming galaxies  (or H {\sc ii}s; $\sim 3\%$).   
For some of these, the maser luminosity is only a smidgen bellow the $10 L_{\sun}$ limit that generally (albeit arbitrarily) defines the megamaser activity, and we have now increasing evidence and reasonable physical explanations for variability in megamaser disks (e.g., Pesce et al. 2015) suggesting that these systems might turn into the next disk detection with new observations. 

Moreover,  if indeed there is an evolutionary sequence in which galaxies transform from star-forming via AGN to quiescent as suggested by several studies (Constantin et al.  2008, 2009, 2015; Schawinski et al. 2007; 2010, Cen 2012), accretion onto the central BH should happen as early as the star formation (or H{\sc\,ii}) phase, and should be active in the Transition Objects.    Thus, if maser activity is related to accretion, we should look for maser disks in H{\sc\,ii}'s and  Transition Objects as well, especially if the evolutionary sequence resembles H{\sc\,ii} $\rightarrow$ Transition Objects $\rightarrow$ Seyfert $\rightarrow$ LINER.   The accreting BH could either be obscured by large amounts of dust  in the surrounding star-forming regions, or simply optically hidden in a mix of other ionization sources (shocks, turbulence, etc.) or host galaxy light.  Hence, {\it restricting our searches to  Seyferts and LINERs  could overlook a good portion, or even the lion's share of megamasers.}   This idea is supported by the detection of a megamaser ostensibly in a star-forming (H{\sc\,ii}) galaxy (i.e., NGC 2989, Braatz \& Gugliucci, 2008), and another one in a galaxy that was not previously identified as an AGN (UGC 3789; Braatz et al. 2010).


\subsection{Megamasers and Their Hosts: X-ray and Radio Studies} \label{limitations_x}

Water maser hosts have been scrutinized outside the optical wavelengths as well, most prominently in X-ray and radio wavelengths.  Such studies offer some important clues to the connection between the water maser emission and the AGN activity, however, they have addressed almost exclusively the maser emission in type 2 Seyferts,  and for rather small number statistics.


To date, 56 masing galaxies that have been investigated in 2-10 keV (Greenhill et al. 2008, Zhang et al. 2006, Zhang et al. 2010, Castangia et al. 2013, Masini et al. 2016), suggest that the masers are found in sources with large absorbing column densities ($N_{\rm H} > 10^{23}$ cm$^{-2}$) at a significance level of 95\%.  Intriguingly however, these studies also find that the available (small) subsamples of nuclear low luminosity (or kilo-) masers ($L_{\rm H_2O} < 10 L_{\sun}$) and disk megamasers show average neutral hydrogen column densities that are indistinguishable from those of the entire megamaser sample and from, e.g., samples of narrow-line Seyfert galaxies that were not selected with respect to maser emission.
Potentially, a clumpy cloud structure in the circumnuclear environment, diverging positions between maser and nuclear sources, and/or an occasional amplification of a background radio continuum source are sufficiently decoupling X-ray--measured column densities and \h2o ~maser properties to mask a clear correlation.    It is also possible that any observed association between \h2o ~megamaser emission and high X-ray column may be subject to selection effects: there are only a handful of galaxies with availability of both quality maser maps and hard X-ray coverage that allow for accurate estimates of $N_{\rm H} > 10^{24}$ cm$^{-2}$  values, and thus for which it is possible to establish meaningful connections between the properties of the masing disk and those of the obscuring torus;  they are also among the nearest of the sample of galaxies searched for maser activity, and do not comprise a complete sample by any means (e.g., Masini et al. 2016).


Radio investigations of maser host galaxies appear to bring additional support to the importance of the AGN activity in the centers of these galaxies.  Comparisons of the radio emission in maser hosts and a control sample of Seyfert 2s show that the maser galaxies exhibit multiband radio luminosities that are higher than those of non-masers by a factor of a few, however, there seems to be no statistical difference in their spectral indices (Zhang et al. 2012; Liu et al. 2017).  If nuclear radio continuum luminosity is an indicator of AGN power (Diamond-Stanic et al. 2009), these findings suggest that  AGN activity and its strength may play an important role in exciting strong H$_{2}$O megamaser emissions.



Nevertheless, these studies do not answer the question of whether water megamasers are always produced in association with an AGN, or whether the maser emission is associated with unrecognized AGN activity (i.e., low-luminosity or obscured AGN) remains unclear.  Also, the possibility that the megamaser activity is connected to other nuclear properties of their host galaxies remains relatively unexplored.  \\

\subsection{The Power of \WISE\ in Identifying New Megamasers} \label{subsec:whywise}

Assuming that all megamaser disks are associated with SMBH accretion, we would need to understand why some types of AGNs are more prone to hosting maser disks than others, even within those classified as type 2.   
Because hot dust surrounding AGNs produces a strong mid-infrared continuum and infrared spectral energy distribution (SED) that is clearly distinguishable from normal star-forming galaxies for both obscured and unobscured AGNs in galaxies where the emission from the AGN dominates over the host galaxy emission, mid-infrared observations offer reliable diagnostics for discovering elusive AGNs (e.g. Lacy et al. 2004; Stern et al. 2005; Alonso-Herrero et al. 2006; Donley et al. 2007; Stern et al. 2012; Assef et al. 2013).
Thus, by employing a systematic study of the mid-IR properties of known maser and non-maser galaxies we are bound to constrain the means by which the obscuring torus and the masing disk are related.   
By investigating how the dust properties compare among the different maser strengths and morphologies we might be able to find out whether a dusty torus is necessary for strong maser disk emission.  Would they reflect similar or different temperature ranges, and thus different mid-IR spectral energy distributions (SEDs) and colors? Would any similarities or differences in their mid-IR characteristics help us build more efficient maser disk surveys?

The all-sky survey carried out by the \emph{Wide-field Infrared Survey Explorer (WISE};  Wright et al. 2010) has opened up a new window in the search for optically hidden AGNs in a large number of galaxies. 
In particular, for low redshift galaxies, the $W1 (3.4 \mu$m) -- $W2 (4.6 \mu$m) color is considerably redder ($\ga 0.8$) in objects where emission is dominated by AGNs than that of inactive galaxies (see Figure 1 in Stern et al. 2012; Assef et al. 2013).  While this criterion alone would likely identify heavily obscured Compton-thick sources that are expected to be good places to search for H$_{2}$O megamasers (Zhang et al. 2006; Greenhill et al. 2008; Castangia et al. 2013), it loses its reliability at fainter fluxes (Assef et al. 2013).  Lower luminosity AGNs (e.g. with Eddington ratios $\la$10$^{-2}$) where strong disk maser systems often reside (Constantin 2012) can be easily missed by this simple selection criterion (Gao et al. 2017).  We are exploring here for the first time a wide variety of the \WISE\ mid-IR photometric properties that could be used as megamaser diagnostics.

Adding information from optical bands proved to increase the success of the mid-IR photometric searches for elusive low luminosity AGNs (e.g. Fiore et al. 2008; Georgantopoulos et al. 2008; Rovilos et al. 2011), and we also probe here these effects towards increasing the megamaser detection rates as well.   In particular, an enhanced 24 $\micron$-to-optical flux ratio implies strong dust heating by the nuclear accreting power source which is optically blocked by the obscuring dust.   
Thus, high 24 $\micron$-to-optical flux ratios (and red $R-K$ colors) can be effective indicators of highly obscured Compton-thick AGNs (e.g., Fiore et al. 2008).   Moreover, because the 24 $\micron$-to-optical flux ratio correlates with the X ray--to--optical ratio in AGNs,  it can act as an estimator of the relative level of AGN activity, and possibly be used to identify low-luminosity and moderately obscured ($N_{H}$ $\sim$8$\times$10$^{22}$ cm$^{-2}$) AGNs (e.g., Georgantopoulos et al. 2008), and thus, megamasers.   

In this paper, we investigate the dependences of the H$_{2}$O megamaser emission and its properties on the global optical/mid-infrared properties of their host galaxies.  Identifying such correlations should help us better understand the underlying maser excitation physics while providing new guidance towards optimizing the sample selection criteria for future maser surveys. In section 2, we describe the samples of maser and non-maser galaxies, and how we collect the associated optical/mid-infrared photometric information.
We present the general optical and mid-IR properties of our samples and the correlations with the various types of maser emission in Section 3, and we explore the physics of the association of the megamaser emission with obscured AGN activity in Section 4.   We summarize the key points of this study and provide guidance for future maser surveys in Section 5.

Throughout this paper we adopt a $\Lambda$CDM cosmology, where $\Lambda$ denotes the cosmological constant that accounts for dark energy, and the universe contains cold dark matter (CDM).   We adopt $\Omega_m = 0.3$, $\Omega_{\Lambda}= 0.7$, and \ho $= 70$ km s$^{-1}$ Mpc$^{-1}$ for the cosmological parameters.

\section{The Maser and Non-maser Galaxy Samples and the Optical/mid-IR Photometric Data}

\subsection{The Sample} \label{sample}


The largest and most comprehensive catalog of all galaxies surveyed for water maser emission at 22 GHz is being compiled and made publicly available by the Megamaser Cosmology Project (MCP; Braatz et al. 2010).   Starting in 2005, MCP conducted maser searches using the \emph{Green Bank Telescope (GBT)} with 1 $\sigma$ rms sensitivity of $\sim 2-3$ mJy per 24.4 kHz ($\sim$0.33 km~s$^{-1}$) channel (e.g. Braatz et al. 2004, Braatz \& Gugliucci 2008). The redshifts of the galaxies in the \emph{GBT} surveys range from $z =0.0 - 0.07$.   The total number of galaxies listed by the MCP website by September 2016 is 4863 galaxies.   Among all the galaxies catalogued by MCP, maser emission is found in 161 galaxies and the overall maser detection rate is therefore $\sim 3\%$.   

We note that while we refer to the data listed on the MCP site as the ``MCP samples or catalogs," they include galaxies that are not the result of MCP surveys.  In particular, the following 19 maser galaxies are not the product of MCP searches, nor have been followed-up with \emph{GBT} as part of the MCP surveys: IC10, ESO-013-G012, M33, IC 184, IC 342, NGC 2146, VII Zw073, He 2-10, J1028+1046, NGC 3256, NGC 3620, NGC 4038/39, NGC 4214, ESO 269-G012, NGC 4945, NGC 5253, Circinus, NGC 6300, ESO 103-G35.    We also note that the entire sample of galaxies we work with here reflects a simple annexation of all galactic nuclei surveyed in 22 GHz, 
and that the targeting criteria and sensitivities associated with each constituent survey have not been homogeneous.  As a consequence, the overall sample might be affected by (to date, unknown) selection effects, however, none of the photometric measurements considered in this analysis  played a major role in selecting these galaxies as potential maser targets.



\LongTables
\begin{deluxetable*}{lrrlrr} 
\tablewidth{0 pt} 
\tablecaption{General properties of the  H$_{2}$O maser galaxies 
\label{tbl-prop}} 
\tablehead{ 
\colhead{Galaxy} & \colhead{Distance } & \colhead{$L_{\rm H_{2}O}$}  &  \colhead{Galaxy}  &  \colhead{Distance}   &  \colhead{$L_{\rm H_{2}O}$}  \\  
\colhead{Name}      & \colhead{(Mpc)}   & \colhead{($L_{\odot}$)} & \colhead{Name} & \colhead{(Mpc)} & \colhead{($L_{\odot}$)} 
}     
\startdata 
                    NGC23   &  65.2   & 180.0  &    2MASXJ11033836-0052081   &  122.6   &  26.0    \\ 
                    NGC17   &  84.7   &  9.0  &    2MASXJ11093314+2837393   &  162.3   &  36.0    \\ 
   2MASXJ00114518-0054303   & 205.5   & 527.0  &                   NGC3556   &   10.0   &   1.0    \\ 
   2MASXJ00272528+4544279   & 171.5   & 507.0  &                   NGC3620   &   24.0   &   4.0    \\ 
                     IC10   &   0.7   & 0.02  &               CGCG185-028   &  149.4   &  550.0    \\ 
                  NGC235A   &  95.2   & 41.0  &                   NGC3614   &   33.3   &   3.0    \\ 
                   NGC253   &   3.4   &  0.2  &                    Arp299   &   44.1   &  125.0    \\ 
                   Mrk348   &  64.4   & 400.0  &                   NGC3735   &   38.5   &  20.0    \\ 
                   NGC291   &  81.5   & 74.0  &               CGCG068-013   &  152.9   &  92.0    \\ 
              ESO013-G012   &  72.1   & 500.0  &                   NGC3783   &   41.7   &  22.0    \\ 
2MASXJ01094510-0332329\tablenotemark{a}   & 233.8   & 1138.0  &               CGCG268-089   &  113.2   &  149.0    \\ 
                     Mrk1   &  68.3   & 50.0  &                NGC4038/39   &   23.5   &   8.0    \\ 
                  NGC520a   &  32.4   &  1.0  & 2MASXJ12020465+3519173\tablenotemark{a}   &  146.0   &  520.0    \\ 
2MASXJ01260163-0417564\tablenotemark{a}   &  80.4   & 112.0  &                   UGC7016   &  103.9   &  22.0    \\ 
                      M33   &   0.8   &  0.3  &                   NGC4051   &   10.0   &   2.0    \\ 
  NGC591\tablenotemark{a}   &  65.0   & 25.0  &                   NGC4151   &   14.2   &   0.6    \\ 
                   NGC613   &  21.1   & 20.0  &                   NGC4194   &   35.7   &  12.0    \\ 
                    IC184   &  75.5   & 20.0  &                   NGC4214   &    4.2   &  0.03    \\ 
   2MASXJ02140591-0016371   & 160.1   & 74.0  &                   NGC4253   &   55.4   &   9.0    \\ 
 MRK1029\tablenotemark{a}   & 129.7   & 682.0  &  NGC4258\tablenotemark{a}   &    7.2   &  80.0    \\ 
                  NGC1052   &  21.0   & 125.0  &                   NGC4261   &   32.0   &  40.0    \\ 
 NGC1068\tablenotemark{a}   &  16.2   & 160.0  &                   NGC4293   &   12.7   &   1.0    \\ 
                  NGC1106   &  60.1   &  8.0  &  NGC4388\tablenotemark{a}   &   36.1   &  12.0    \\ 
   2MASXJ02532956-0014052   & 124.2   &  4.0  &                   NGC4527   &   24.8   &   4.0    \\ 
 Mrk1066\tablenotemark{a}   &  51.5   & 32.0  &                   NGC4633   &   22.6   &  0.06    \\ 
 NGC1194\tablenotemark{a}   &  58.2   & 112.0  &    2MASXJ12555984-0804329   &   57.3   &  266.0    \\ 
 NGC1320\tablenotemark{a}   &  38.0   & 43.0  & ESO269-G012\tablenotemark{a}   &   71.6   &  1000.0    \\                  
   2MASXJ03364614-0750236   & 167.5   & 183.0  &                  NGC4922B   &  100.8   &  200.0    \\ 
 NGC1386\tablenotemark{a}   &  12.4   & 125.0  &                   NGC4945   &    8.0   &  50.0    \\ 
IRAS03355+0104\tablenotemark{a}   & 170.4   & 227.0  &  NGC4968\tablenotemark{a}   &   42.2   &  53.0    \\ 
                    IC342   &   3.3   & 0.01  &                   NGC5077   &   40.1   &  71.0    \\ 
               2MFGC03185   & 176.8   & 3870.0  &                   NGC5128   &    7.8   &   1.0    \\ 
WISEPJ043703.69+245606.9\tablenotemark{a}   &  69.1   & 178.0  &                      M51a   &    7.8   &   0.6    \\ 
2MASXJ04370825+6637424\tablenotemark{a}   &  53.9   & 20.0  &                   NGC5256   &  120.3   &  32.0    \\ 
           IRAS04385-0828   &  64.7   & 43.0  &                   NGC5253   &    3.5   &   1.0    \\ 
                  UGC3157   &  66.0   & 85.0  & SBS1344+527\tablenotemark{a}   &  125.2   &  379.0    \\ 
 UGC3193\tablenotemark{a}   &  63.6   & 270.0  &                   NGC5347   &   33.4   &  32.0    \\ 
                  Mrk1089   &  57.4   & 15.0  &    2MASXJ13553592+0553050   &  168.0   &  86.0    \\ 
              CGCG468-002   &  75.0   & 35.0  &             MCG+11-17-010   &  137.0   &  34.0    \\ 
                  UGC3255   &  81.0   & 16.0  &    2MASXJ13564736+6937117   &  150.3   &  397.0    \\ 
                  UGCA116   &  11.3   &  3.0  &               ESO446-G018   &   68.2   &  183.0    \\ 
                     Mrk3   &  57.9   & 10.0  &                   UGC9077   &   80.4   &  17.0    \\ 
                  NGC2146   &  12.8   &  1.0  &  NGC5495\tablenotemark{a}   &   96.2   &  600.0    \\ 
                 VIIZw073   & 177.0   & 137.0  & Circinus\tablenotemark{a}   &    4.2   &  20.0    \\                    
 NGC2273\tablenotemark{a}   &  26.3   & 32.0  &                   NGC5506   &   26.5   &  50.0    \\ 
ESO558-G009\tablenotemark{a}   & 109.6   & 766.0  &             MCG-02-37-004   &  177.7   &  329.0    \\ 
 UGC3789\tablenotemark{a}   &  47.5   & 370.0  &                   NGC5643   &   17.1   &  25.0    \\ 
                  NGC2410   &  64.2   & 14.0  &                   NGC5691   &   26.7   &   4.0    \\ 
   Mrk78\tablenotemark{a}   & 159.1   & 32.0  &  NGC5728\tablenotemark{a}   &   40.0   &  80.0    \\ 
                    IC485   & 119.1   & 1060.0  &               CGCG164-019   &  128.0   &  30.0    \\ 
 Mrk1210\tablenotemark{a}   &  57.8   & 80.0  & NGC5765b\tablenotemark{a}   &  119.0   &  2277.0    \\ 
                   He2-10   &  12.5   &  0.7  & UGC9618b\tablenotemark{a}   &  144.2   &  400.0    \\ 
2MASXJ08362280+3327383\tablenotemark{a}   & 211.6   & 800.0  &  UGC9639\tablenotemark{a}   &  154.3   &  444.0    \\ 
                  NGC2639   &  47.7   & 25.0  &  NGC5793\tablenotemark{a}   &   49.9   &  200.0    \\ 
2MASXJ08474769-0022514\tablenotemark{a}   & 218.2   & 2210.0  &    2MASXJ15201964+5253560   &  159.2   &  44.0    \\ 
              CGCG120-039   & 109.8   & 154.0  &                   UGC9954   &  147.4   &  30.0    \\ 
   2MASXJ09053008+0328232   & 120.3   & 132.0  &    2MASXJ16070391+0106296   &  116.7   &   6.0    \\ 
                  NGC2781   &  29.3   & 14.0  &            IRAS16288+3929   &  131.2   &  16.0    \\ 
   2MASXJ09124641+2304273   & 155.6   & 143.0  &               CGCG168-018   &  157.4   &  192.0    \\ 
                  NGC2782   &  36.6   & 13.0  &                   NGC6240   &  104.8   &  40.0    \\ 
                  NGC2824   &  39.4   & 500.0  &  NGC6264\tablenotemark{a}   &  145.4   &  2100.0    \\ 
              SBS0927+493   & 146.3   & 280.0  & 2MFGC13581\tablenotemark{a}   &  145.6   &  672.0    \\ 
                  UGC5101   & 168.6   & 1900.0  &    2MASXJ17101815+1344058   &  135.0   &  69.0    \\ 
 Mrk1419\tablenotemark{a}   &  70.5   & 400.0  &  NGC6323\tablenotemark{a}   &  111.0   &  500.0    \\ 
                  NGC2979   &  38.9   & 125.0  &                   NGC6300   &   15.8   &   3.0    \\ 
                  NGC2989   &  59.2   & 40.0  &               CGCG252-044   &  105.1   &  110.0    \\ 
                      M82   &   3.7   & 200.0  &                  UGC11035   &  111.4   &  291.0    \\ 
                  NGC3081   &  33.8   & 17.0  &                ESO103-G35   &   56.9   &  400.0    \\ 
 NGC3079\tablenotemark{a}   &  16.1   & 500.0  &    2MASXJ19393889-0124328   &   88.4   &  160.0    \\ 
   2MASXJ10115058-1926436   & 115.1   & 371.0  &                     3C403   &  252.7   &  2000.0    \\ 
                  NGC3160   &  97.9   & 46.0  &  NGC6926\tablenotemark{a}   &   85.3   &  500.0    \\ 
  IC2560\tablenotemark{a}   &  41.4   & 100.0  &            IRAS20550+1656   &  154.6   &  908.0    \\ 
   2MASXJ10205956-0010342   & 239.7   & 65.0  &                  UGC11685   &   83.9   &  80.0    \\ 
                  NGC3256   &  40.1   & 10.0  &                    IC1361   &   56.6   &  35.0    \\ 
               2MFGC08115   & 191.0   & 1135.0  &    2MASXJ21445731+1534503   &  130.5   &  189.0    \\ 
 UGC5713\tablenotemark{a}   &  90.2   & 237.0  &           AM2158-380NED02   &  142.6   &  559.0    \\ 
   MRK34\tablenotemark{a}   & 216.3   & 1000.0  &               TXS2226-184   &  107.1   &  6300.0    \\ 
                  NGC3359   &  14.5   &  0.7  &                   NGC7479   &   34.0   &  19.0    \\ 
 NGC3393\tablenotemark{a}   &  53.6   & 320.0  &   IC1481\tablenotemark{a}   &   87.4   &  320.0    \\ 
 UGC6093\tablenotemark{a}   & 154.7   & 770.0  &             MCG+05-55-041   &  133.7   &  93.0    \\ 
 CGCG498-038\tablenotemark{a} &  132.0 &  268.0 &                            &          &          \\
\enddata 
\tablecomments{  
Columns 1 \&  4: Name of the galaxy; Columns 2 \& 5: The adopted distance of the galaxy (see text); Columns 3 \& 6: The isotropic H$_{2}$O maser luminosity.  Objects are listed in order of their right ascension.   
\tablenotetext{a}{maser emission in a disk-like configuration; see https://safe.nrao.edu/wiki/bin/view/Main/MegamaserCosmologyProject for more detail about these maser detections.}}
\end{deluxetable*} 

The data presented in the MCP catalogs comprise sky positions, recession velocities, 22 GHz spectra, the sensitivity of each observation, and the corresponding source brightness temperature.  Herein, we refer to galaxies with confirmed H$_2$O maser emission as maser galaxies or masers, and to galaxies with no maser detection as non-maser galaxies, or non-masers. For the maser galaxies, the MCP catalogs also include the maser luminosities, morphologies (e.g., disk, jet), and the corresponding discovery reference.   For those galaxies without available $L_{\rm H_{2}O}$ from the MCP website, we calculate the values based on the \emph{GBT} spectra.   The maser luminosities $L_{\rm H_{2}O}$ are calculated under the assumption of isotropic emission (see Pesce et al. 2015 for the definition).   
The MCP directory includes also the only two high redshift galaxies with maser detection: J0804+3607 and J0414+0534 (z=0.66 and z=2.64, respectively).  These two sources have strikingly different properties, both in terms of their hosts and in their maser emission, and we do not include them in our analysis.          
Table \ref{tbl-prop} lists the general properties of the 161 low-$z$ maser galaxies, including their distances and their water maser luminosities.     
The distances are estimated by dividing the recession velocities of the galaxies ($v_{\rm sys}$) by \ho, when $v_{\rm sys} \geq 600$ km/s; otherwise, in order to avoid significant distance bias caused by peculiar velocities, we recorded and used here the redshift independent distances from NED\footnote{https://ned.ipac.caltech.edu}.


\subsection{The Classification of Maser Types} \label{maserclass}

\begin{figure*}[ht]
\begin{center} 
\vspace*{0 cm} 
\hspace*{-1 cm} 
\includegraphics[angle=0, scale=0.47]{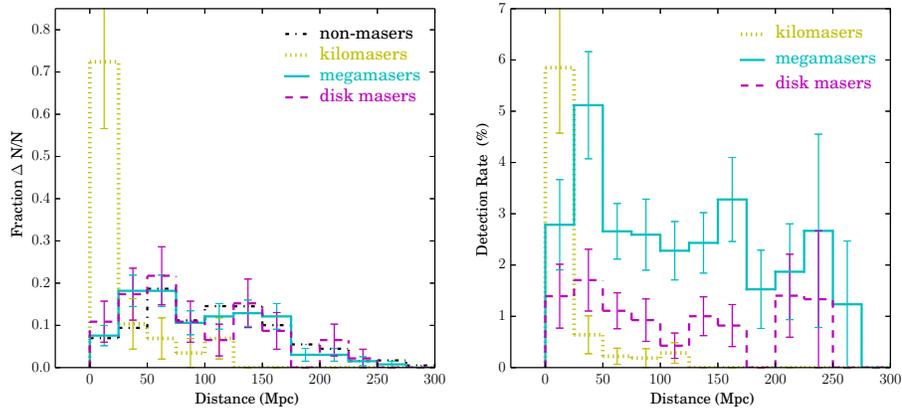}
\vspace*{0.0 cm} 
\caption{{\bf Left panel:}  Normalized number of masers and non-masers in the MCP sample as functions of distance. {\bf Right panel:} The maser detection rates as a function of distance.  The error bars illustrate the associated Poisson uncertainties. }
\label{mstat}
\end{center} 
\end{figure*}

For the sample of galaxies with maser detections, we separate them into two general categories based on their isotropic maser luminosities $L_{\rm H_{2}O}$.  The low-luminosity category is called ``kilomasers", which comprises 28 galaxies with $L_{\rm H_{2}O}$ $<10$  $L_{\odot}$.  The kilomaser emission is more likely to be associated with starforming activity rather than an AGN.  The higher luminosity category is called "megamasers", which includes 132 galaxies with $L_{\rm H_{2}O}$ $\geq$ 10 $L_{\odot}$.  Most of the H$_{2}$O megamaser activity is believed to be associated with AGNs, however starburst activity cannot be ruled out (e.g., the optical spectral classification of $\sim 25$\% of these galaxy nuclei are consistent with either H {\sc ii}s or Transition Objects; Constantin 2012).

Within the category of megamasers, we further define a subsample called ``disk masers" of objects where the H$_{2}$O maser emission exhibits the characteristic spectral pattern of high-velocity features that are  redshifted and blueshifted and spatially offset from the features of the systemic velocity (e.g. Herrnstein et al. 1999; Braatz \& Gugliucci 2008), or are identified as such from direct mapping of the spatial distribution of the maser emissions using the VLBI technique (e.g. Kuo et al. 2016, Gao et al. 2016).    
In all of these cases, it is found that the water maser emission originates from sub-parsec scale circumnuclear gas disks at the centers of galaxies (e.g. Herrnstein et al. 1999, Kuo et al. 2011; Gao et al. 2016). This is the only type of maser that allows accurate measurements of BH masses and of the Hubble constant (e.g Reid et al. 2013, Kuo et al. 2013, Kuo et al. 2015, Gao et al. 2016) and is therefore the type of maser system we are most interested in detecting more of.     
The total number of disk masers identified by this method is 45 and the overall detection rate of such systems is 0.9\%.   
The remaining megamasers are likely to be associated with outflows or jets, or disk masers with blueshifted/ redshifted masers which are too weak to be detected given the sensitivity of our survey;  however, the scarce data availability on these cases does not allow us to quantify the real fraction of these situations. 
 
Figure~\ref{mstat} presents a summary of the number statistics of the samples of maser and non-maser galaxies as a function of distance. The left panel shows the normalized distributions of non-masers (dot-dashed black), masers (dotted yellow), and the sub-groups of megamasers (dashed magenta) and disks (continue cyan). $\Delta$N is the number of objects in a given sample in a distance bin, while N is the size of the sample. The right panel shows the detection rates of masers, megamasers, and disks, out of all the galaxies surveyed for water maser emission, as a function of distance.   
As it is readily apparent from these distributions, the majority of the maser galaxies have been detected within 150 Mpc, and the detection rate decreases with increasing distance. Nevertheless, at distances larger than 50 Mpc, the statistics are not high enough to determine with confidence or quantify the distance dependence.

\subsection{Uncertainties in the Maser Detection Rates}

The megamaser searches are likely to suffer from variations of the detection limits in terms of their luminosity, because their detection is sensitive not just to the presence of a maser system but also to the flux densities of the individual maser lines.   More specifically, detection limits for maser surveys are generally cited in terms of the root-mean-square flux density per spectral channel, and thus, the sensitivity can be restated in terms of the isotropic maser luminosity for a given line width.   Since a maser system may consist of many isolated maser components, it is possible that in the case of a maser system with multiple features for which no single channel exceeds the detection threshold, the maser system may remain undetected.   Thus, a maser survey with a uniform luminosity limit is not feasible,  and there is no practical way to estimate the number of missing maser systems in a given distance (or other parameter) range.   As a consequence, the errors associated with the maser detection rates cannot account for this effect. 


We also want to emphasize that what we call ``detection rate" is simply defined as the fraction of objects that are detected in 22 GHz emission to all of the objects surveyed for this kind of emission, in the respective bin of host properties, and therefore it simply reflects the number of all the systems above the detection threshold, regardless of how luminous these sources are in their 22 GHz activity, or in what cosmic volume they have been detected.  We  point out that we do not (and are not yet able to) estimate or calculate the true prevalence of maser emission in host galaxies of certain properties. 

On the other hand, there are additional uncertainties that could particularly affect histogram comparisons (e.g., Figure~\ref{mstat}).    Specifically, the values we list and compare in (any) histograms are subject to uncertainties in the values that are being binned up to create the histogram (i.e., the Òbin hoppingÓ effects).    We checked, however, these effects on the validity of the conclusions based on histogram comparisons throughout this study, by employing  comparisons of histograms built by bootstrapping, i.e., adding/subtracting objects from various bins and placing them in the adjacent ones, and we did not identify any significant change.  In particular,  using a simple Monte Carlo (MC) approach, we create the same histogram many times, each time replacing each data point by a new one drawn from the distribution corresponding to that data point's uncertainty (i.e., a Gaussian with a mean equal to the data point's value and variance equal to the square of its uncertainty).Ê By generating a large number of these histograms and looking at the root-mean-square of the value within any particular bin, we are able to get a very robust estimate of the ``bin hopping" uncertainty allowed by the dataset.Ê Such a technique has the added benefit of also determining the ``most likely" value for the histogram bin itself, by averaging the values we get from the MC process.Ê We find that the result of adding uncertainty to the measurements is to: (1) smooth out the overall shape of this histogram, and (2) increase the value of the error bar on any single bin.Ê However, for this particular sample and its subgroups, the Poisson uncertainties remain dominant.

\subsection{Optical Photometry of the Galaxy Samples} \label{opticaldata}

The optical photometric data for our MCP samples come primarily from the \emph{Sloan Digital Sky Survey (SDSS}; York et al., 2000; Abazajian et al., 2009; Ahn et al., 2014), and we adopt the SDSS $u$, $g$, $r$, $i$, $z$ magnitude values reported in NASA/IPAC Extragalactic Database (NED).  Out of the total MCP sample of 4863 galaxies, NED provides SDSS magnitudes for 2753 galaxies (i.e., 57\% of the sample).   Among these galaxies, 2731 of them have various types SDSS magnitudes coming from different model fittings (i.e. PSF, Model, CModel, Petrosian\footnote{see http://www.sdss.org/dr12/algorithms/magnitudes/}).  We adopt here the ``Model" magnitudes, which provide good estimates of global galaxy fluxes independent of aperture sizes.  For the rest of the 32 nearby galaxies where ``Model" photometry is not available (the SDSS catalog is missing many nearby, bright galaxies because of the difficulty of automatic photometric processing of big galaxies) NED provides SDSS photometry from observations with aperture sizes large enough 
to cover the bulk of the galaxy except for seven systems (NGC 3521, NGC 3627, NGC 4631, NGC 4826, F15327+2340, NGC 6269, and NGC 7331).  We exclude these latter galaxies from the statistical analysis that involves optical photometric properties.  

Among the 2749 galaxies with available SDSS photometry, there are 63 megamaser detections, among which 18 show a disk-like configuration.   Thus, in terms of optical photometry, the overall detection rate of all masers, megamasers and disk masers are therefore 2.3\%, 1.9\% and 0.7\%, respectively. These values are slightly lower than those of the whole MCP sample, however, in the same ratio, implying that selection biases that generally affect the various maser type detection efficiency should not be significantly different for the subset  of galaxies with SDSS photometry. Nonetheless, any comparisons between the entire MCP and the SDSS-MCP samples should be made with some caution because the two samples may have some intrinsic differences in terms of galactic properties. 




\subsection{Mid-Infrared Photometry of the Galaxy Samples} \label{wisedata}

We collect the mid-infrared photometric data of our MCP samples entirely from the public all-sky source catalog provided by the  \emph{Wide-field Infrared Survey Explorer (WISE}; Wright et al. 2010) via NASA/IPAC Infrared Science Archive\footnote{http://wise2.ipac.caltech.edu/docs/release/allsky/} using the IRSA\footnote{http://irsa.ipac.caltech.edu/Missions/wise.html} catalog tool.   \emph{WISE} mapped the sky at 3.4, 4.6, 12, and 22 $\mu$m (W1, W2, W3, W4) in 2010 with an angular resolution of 6.1\arcsec, 6.4\arcsec, 6.5\arcsec, and 12.0\arcsec\ in these four bands, respectively.

Our cross-matching technique searches for all \emph{WISE} detections within an initial cone search radius of 2\arcsec, with a gradual increase of the search radii of up to 10\arcsec, with a step size of 0.5\arcsec.   For comparison, the \emph{GBT} beam size at 22 GHz is $\sim 30$\arcsec. We require for all the MCP-\emph{WISE} matches to be positioned within 10\arcsec\ from the \emph{GBT} source positions, and to be detected at better than 3$\sigma$ in both the 3.4$\mu$m and 4.6$\mu$m bands (W1 and W2, respectively), which are used to identify and classify these galaxies as AGN based on the W1--W2 color (see Section ~\ref{subsec:wisecolor}, Figure~\ref{Wcolors}).  We also require for the matches to have signal-to-noise ratios of SNR $\geq 3$ in the W4 band. We did not restrict the detection SNR in the W3 band because galaxies that meet our requirements in W4, which is the least sensitive band, will generally have higher signal-to-noise detections in the W3 band.   For all duplicates and search radii larger than 6\arcsec, we visually inspected the \WISE\ matches in all four \emph{WISE} bands to make sure that it corresponds to the MCP source.  The proximity of all these galaxies allows unambiguous detections for the great majority of our sources (86\% of the entire sample; 4174 galaxies) in all four passbands;  all of the kilomasers and the megamaser disks are found to have a \emph{WISE} counterpart while the megamasers only have a 98\% \WISE\ match.  

Overall, the overlap with \emph{WISE} is higher than with the optical photometry, allowing for a greater statistical significance of the investigation of the mid-IR features that may link to the masing activity in galaxy centers.   The high MCP-\WISE\ matching rate for the megamasers and disks is somewhat expected, if not encouraging, since these systems are expected to have strong mid-IR emission, and therefore highly detectable counterparts given the previously proposed scenarios for maser emission associated with the circumnuclear dusty torus in AGNs, which (re)emits in the mid-IR regime.   Nevertheless, accounting for the maser selection bias (as being targeted out of Seyfert galaxies; Constantin 2012), it is important to investigate whether the maser-Seyferts differ from other Seyferts in their mid-IR emissions due to properties related to the maser emission.

\section{Results: The Observed Correlations}

\subsection{Optical Photometric Properties of Maser Galaxies} \label{subsec:optical}

Optical photometric measurements have been relatively successfully employed in the past for mapping the distribution of galaxy properties that can be used to constrain semi-analytic models of galaxy formation evolution processes.   Here we briefly address the degree to which the presence and strength of maser emission in galaxy centers relates to the optical colors and absolute brightness of their hosts.   

\begin{figure*}[ht]
\begin{center} 
\vspace*{0 cm} 
\hspace*{-1 cm} 
\includegraphics[angle=0, scale=0.47]{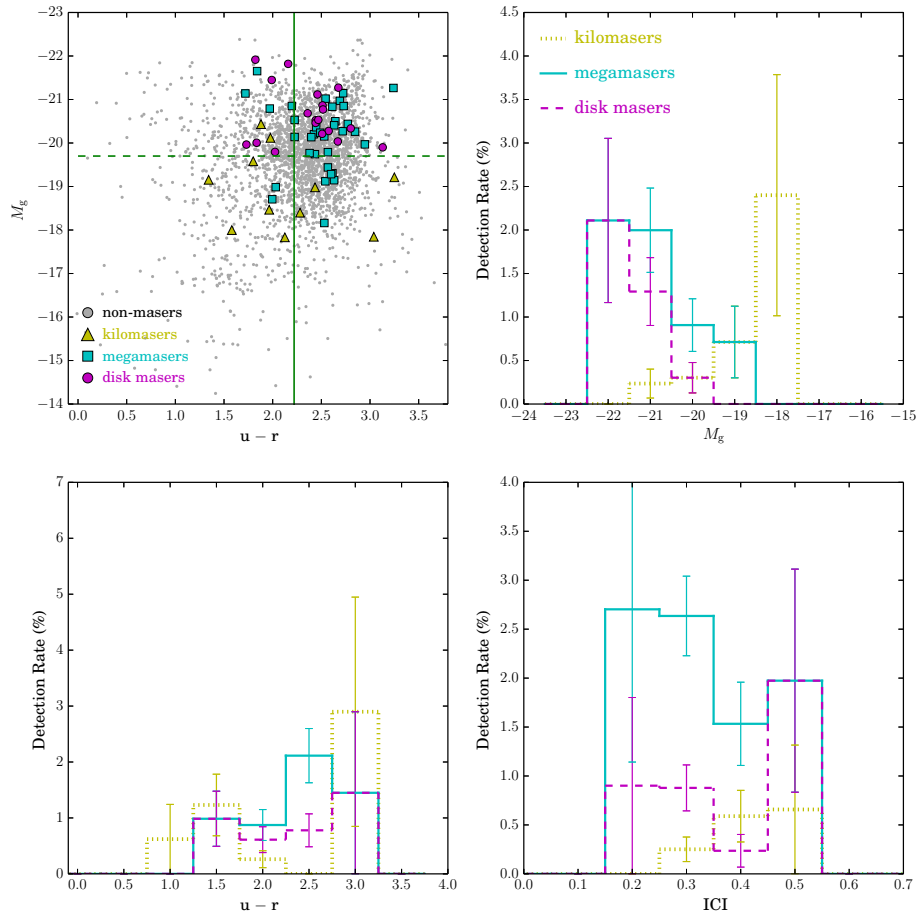}
\vspace*{0.0 cm} 
\caption{{\bf Upper left panel:} The distribution of the sample galaxies in the absolute $g$-magnitude ($M_{\rm g}$) and the $u-r$ color.  The kilomasers are shown as filled yellow triangles, megamasers as cyan squares, disks as magenta filled circles, and non-masers as small grey circles.  The horizontal and vertical dashed lines corresponds to $M_{\rm g} = -19.7$  and $u-r = 2.2$, respectively. {\bf Upper right panel:}The maser detection rates as a function of $M_{\rm g}$.  The kilomasers, megamasers, and disk masers, are shown with the dotted yellow, solid cyan, and dashed magenta lines respectively. The errors correspond to the Poisson uncertainties of the maser detection rates in each bin. {\bf Lower left panel:} The maser detection rates as a function of the $u-r$ colors; $u-r > 2.22$ suggests early type galaxies (E, S0, and Sa) while $u-r < 2.22$ suggests late type galaxies (Sb, Sc, and Sd).   {\bf Lower right panel:}  The maser detection rates as a function of the inverse concentration index (ICI).}
\label{optical}
\end{center} 
\end{figure*}

To compare the colors of maser and non-maser galaxies, we consider the $u-r$ color, which is sensitive to extinction, as well as to the UV flux and the 4000\AA\ break, and thus relates to the age and properties of the associated stellar populations. 
For the absolute brightness we are considering the $g-$band absolute magnitude $M_g$, which is converted from the observed SDSS $g$-magnitude.  We note that the trends found here with  $M_{\rm g}$ are consistent with those exhibited when other SDSS magnitude bands are used.  To compute colors, we use extinction (Schlegel et al. 1998) and K-corrected model magnitudes.  

For a very basic comparison of their morphological properties, we examine the distribution of inverse concentration indices (ICI) measured by the SDSS photometric pipeline (Lupton et al. 2001; Stoughton et al. 2002; Pier et al. 2003).  ICI is defined by the ratio ICI = $r_{50}/r_{90}$, where $r_{50}$ and $r_{90}$ correspond to the radii at which the integrated fluxes are equal to 50\% and 90\% of the Petrosian flux, respectively.  A small value of ICI corresponds to a relatively diffuse galaxy and a large value of ICI to a highly concentrated galaxy.  The concentration index (or the inverse concentration index) has been shown to correlate well with galaxy type (Strateva et al. 2001; Shimasaku et al. 2001).


We show in Figure~\ref{optical} the distribution of the MCP galaxy samples in the $u-r$ color and $M_{\rm g}$, together with the maser detection rates for each subcategory of maser (kilomasers, megamasers, and disk masers) as a function of these two parameters, and as a function of ICI. In the color-magnitude diagram, the yellow triangles, cyan squares, and magenta filled circles  represent the kilomasers, megamasers, and disk masers, respectively; the sample of non-masers are shown in grey bullets.  We are using the same symbol and color scheme to label different types of masers throughout the rest of the paper.  

These plots show that different types of maser galaxies have substantially different distribution in the color-magnitude diagram. In particular, the megamasers tend to be brighter than the kilomasers: 83\% of megamasers lie above $M_{\rm g}$ $=$ $-19.7$ while only 18\% of the kilomasers exhibit this level of optical brightness.   The disk megamasers reside 100\% in these optically brighter galaxies as they all have $M_{\rm g}$ $<$ $-19.7$.  These trends are consistent with the findings of Zhu et al. (2011) who considered $B$-band absolute magnitudes derived from SDSS photometry.

The preference of megamasers and disk masers to lie in galaxies with $M_{\rm g}$ $< -19.7$ is  interesting in light of the findings of Kauffmann et al. (2003) who show that among nearby galaxies, the $M_{\rm g}$ $\sim$ $-20$ magnitude (corresponding to the stellar mass $\sim$3$\times$10$^{10}$ $M_{\odot}$) marks a transition point beyond which there is a rapid increase in the fraction of galaxies with old stellar populations, high surface stellar densities, and high stellar concentrations typical of bulges.   The trends we find here  suggest then that the megamasers prefer optically bright hosts with likely higher stellar masses.   A direct comparison of the dependence of the maser emission properties on stellar characteristics indicators like those used by Kauffmann et al. (2003) is however only possible for a small fraction of our sample (44/162 maser galaxies have SDSS spectroscopy), and preliminary analysis reveals a possible trend towards more massive (and younger) stellar populations, however, at a weak statistical significance (e.g., Constantin 2012).  

In terms of the maser detection rates, we see a clear increase with brighter (lower) $M_{\rm g}$ values for the disks and possibly for the whole subsample of megamasers, whereas the trend may be opposite for the kilomasers.   Nevertheless, despite the fact that  the detection rate of megamasers is enhanced at lower $M_{\rm g}$ up to $\sim$4\%, the improvement based on the optical photometry alone is too small to be useful for future maser survey.  Nonetheless, as we will show in Section ~\ref{subsec:gmag}, when used in conjunction with mid-IR photometric information, the optical photometry becomes quite useful for producing a significant increase in the maser detection.

On the other hand, the maser detection rate of different maser types  reveals a rather weak dependence on the galaxy color suggesting that there is no preference for megamasers to reside in a  particular type of galaxy; while the disks appear to occupy the $u-r > 2.22$ locus, there is  a small fraction of megamasers (1.3\%) with bluer colors.   Given that the $u-r = 2.22$ color can be used to  separate between early (E, S0, and Sa) and late type (Sb, Sc, and Sd) galaxies (Strateva et al. 2001), and that the majority of our maser galaxies with $u-r > 2.22$ are early-type spirals rather than ellipticals, it is possible that our conclusions apply to spiral hosts only.   On a similar note, the ICI parameter does not prove to be useful in delineating megamasers or disks, as their detection rates does not show any particular enhancement for any range;  the kilomasers, however, do appear to prefer galaxies with higher ICI values, which, as expected in these cases, are associated with late type hosts.




\subsection{Mid-infrared Properties of Maser Galaxies} \label{wiseprop}

An association of the maser activity with AGN emission is plausible, however, it is by no means clear how the obscuring torus and the masing disk are related.   
We do not know yet whether megamaser disks are always related to black hole accretion, or whether they require a dusty/molecular torus.  If this torus is necessary for strong maser disk emission, would the properties of the dusty torus be similar or different among all megamaser disks?  Would they reflect similar or different temperature ranges, and thus different mid-IR spectral energy distributions (SEDs) and colors? Would any similarities or differences in their mid-IR SEDs help us build more efficient maser disk surveys? We explore these connections in the following subsections.

\subsubsection{ $W1-W2$ versus $W2-W3$} \label{subsec:wisecolor}

\begin{figure*}[ht] 
\begin{center} 
\vspace*{0 cm} 
\hspace*{-1 cm} 
\includegraphics[angle=0, scale=0.47]{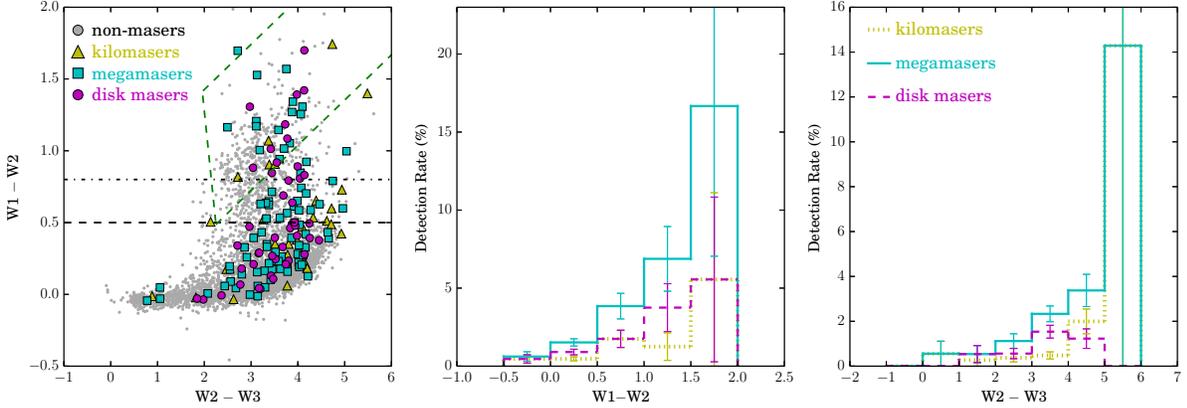}
\vspace*{0.0 cm} 
\caption{{\bf Left panel:} The distribution of the MCP samples of maser and non-maser galaxies in the $W2-W3$ versus $W1-W2$ diagram. The horizontal dashed and dot-dahsed lines correspond to $W1-W2 = 0.5$ and $W1-W2 = 0.8$ cuts, respectively. The green dashed line illustrates  the wedge defined by Mateos et al. (2012); {\bf Middle \& Right panels:} The maser detection rate as a function of $W1-W2$ and $W2-W3$. }
\label{Wcolors}
\end{center} 
\end{figure*}

Previous studies found that we can detect a large portion of the population of red/dusty AGNs that do not show AGN-like optical signatures by using the \WISE\ $W1-W2$ vs. $W2-W3$ color diagrams.  
Mid-IR color cuts proposed for selecting red/obscured AGNs include $W1-W2\geq0.5$ (Ashby et al. 2009), $W1-W2\geq0.8$ for a lower contamination by non-active galaxies (Stern et al. 2012), and sets of criteria involving both of these \emph{WISE} colors, with even more stringent control for non-AGN contamination (e.g., Jarrett et al. 2011; Mateos et al. 2012).   
Figure~\ref{Wcolors} illustrates in the left panel the location of all of our MCP objects in the ($W1-W2$, $W2-W3$) color-color space, along with these \emph{WISE}--AGN cuts.   We chose to work with the Mateos et al. (2012) wedge cut here, and note that the results are very similar if the Jarrett et al. (2011) box is used instead.   


If there is a strong association between megamaser and disk emission and obscured AGN activity, then we expect to see most of these systems with \WISE\ colors of $W1-W2 \geq 0.8$ or within the Mateos et al. (2012) wedge, with most of the non-masers falling below the $W1-W2 = 0.5$ line.  It seems, however, that this is not immediately the case: the masers and non-masers occupy color-color regions that overlap considerably. The distributions of both maser and non-maser galaxies follow the expected spread of galaxies (Wright et al. 2010), without a clear delineation of \emph{WISE} galaxy colors that can be associated with the masers. The only possible difference is in their average $W2-W3$ color, which is clearly redder for the masers ($< W2-W3>$ is $3.51\pm0.06$ for all masers and $3.07\pm0.01$ for non-masers).  Only 24\% and 42\% of the maser galaxies satisfy the criteria $W1-W2\geq0.8$ and $W1-W2 > 0.5$, respectively, while only 18\% of the maser galaxies fall within the color wedge defined by Mateos et al. (2012).   Among megamasers and disks, only $\sim1/3$ fall into the red \WISE\ AGN category, regardless of which criteria are used, which implies that the majority of the disk megamasers are not necessarily associated with obscured dominant AGN activity.  

Interestingly, while these mid-IR criteria for selecting obscured AGN do not seem to be particularly useful in finding a large fraction of megamasers and disks,  the overall maser detection rates skyrocket in the red \WISE\ AGN regime.   We show the maser, megamaser, and disk detection rates as a function of these two colors in the central and the right panels of Figure  ~\ref{Wcolors}.  The overall maser emission is detected among $\sim$ 9\% of these sources, with a high proportion of them being megamasers and disks; there is a significant rise in the expected megamaser and disk detection ($7-9$\% and $2-3$\%, respectively, depending on the color criteria), which represents a boost by at least a factor of three from the currently low rate of $<$3\% and $<$1\%, respectively.    Thus, these trends also suggest that red AGNs are more likely to host a powerful water maser than the bluer AGNs.


Thus, while within the whole population of nearby galaxies there are very few objects with red $W1-W2$ colors ($\sim$4\%; Stern et al.\ 2012), it is definitely of great value to study these systems carefully, and definitely survey them all for maser emission. However, it is also clear that we will only recover  about a third of the whole population of megamasers within this color regime.  It is thus of interest to try to understand which other properties could be used to distinguish between masers and non-masers, especially within the \emph{WISE}--blue galaxy population as well.


\subsubsection{The Role of the Mid-IR Absolute Brightness} \label{lmid_ir}

\begin{figure*}[ht] 
\begin{center} 
\vspace*{0 cm} 
\hspace*{-1 cm} 
\includegraphics[angle=0, scale=0.47]{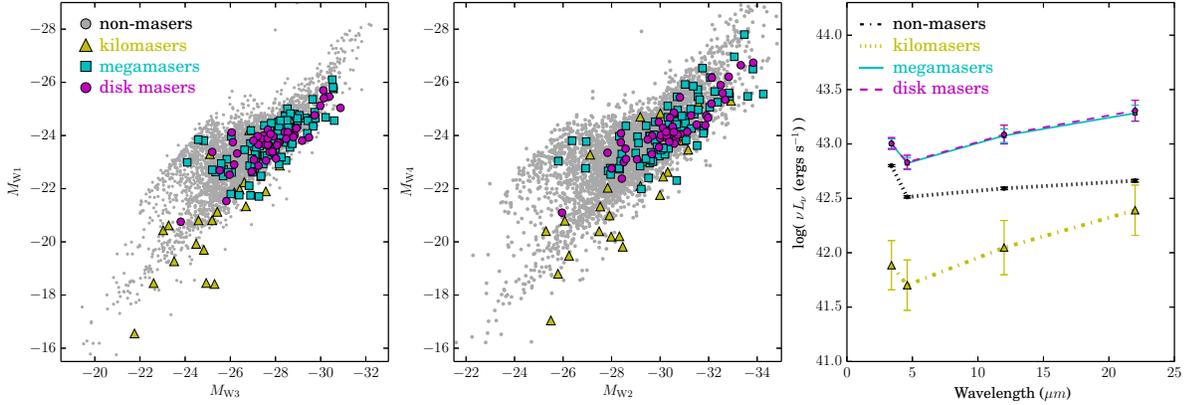}
\vspace*{0.0 cm} 
\caption{ {\bf Left \& Middle panels:} The distribution of the MCP galaxies in the ($M_{W3}$, $M_{W1}$) and the ($M_{W2}$, $M_{W4}$) parameter space.   $M_{W1}$, $M_{W2}$, $M_{W3}$  and $M_{W4}$ are the absolute magnitudes of the \emph{WISE}  bands. {\bf Right panel:} Average $\nu L_{\nu}$ values as a function of the \emph{WISE} mid-infrared wavelengths for the MCP kilomasers (yellow triangles), megamasers (cyan squares), disks (magenta filled circles), and non-masers (small grey circles).}
\label{absmag}
\end{center} 
\end{figure*}

After studying the role of the red AGN--sensitive $W1-W2$ and $W2-W3$ colors in detecting maser galaxies and their subtypes of megamasers and disks, we extend our exploration to all possible mid-IR information that can be provided by the  \emph{WISE} observations.   Among all of the available mid-IR information for these galaxies, we find that the absolute mid-infrared magnitudes ($M_{W1}$, $M_{W2}$, $M_{W3}$, $M_{W4}$) convey important messages for enhancing the water maser detection rates.   We discuss here the most significant trends.

Figure ~\ref{absmag} shows the distribution of our sample of maser and non-maser galaxies in the ($M_{W1}, M_{W3}$)  and ($M_{W2}, M_{W4}$) diagrams, as well as the average mid-IR spectral energy distributions (SEDs; average $\nu$L$_{\nu}$ values calculated for each \emph{WISE} band, plotted as a function of each band's central wavelength) per maser type.     The \emph{WISE} fluxes for each galaxy have been calculated using the magnitude to flux conversion method described in Wright et al.\ (2010), and the luminosities were calculated using the distances listed in Table ~\ref{tbl-prop}.  

The most prominent trend that this comparison reveals is that the megamasers are significantly brighter than both the non-masers and the kilomasers in all of the four \WISE\ bands.   In fact, the mid-IR absolute luminosities or absolute magnitudes appear to be  an efficient separator between the kilomasers and the megamasers, with the latter type being, on average, at least one order of magnitude more luminous (see right panel of Figure  ~\ref{absmag}).   Galaxies without detected maser emission exhibit average absolute luminosities (or magnitudes) in between those of kilomasers and megamasers.    When the  average integrated mid-IR luminosities $L_{\rm MIR}$ are calculated, 
the trends remain equally obvious: $<L_{\rm MIR}>_{\rm kilomasers} = 1.7 \times 10^{43}$ erg $^{-1}$ while $<L_{\rm MIR}>_{\rm megamasers} = 1.4 \times 10^{44}$ erg $^{-1}$.   The megamasers and their subgroup of disks remain similar in their mid-IR total power output, however, it is important to note that the object with the highest luminosity in each of the four \emph{WISE} bands is a disk maser galaxy (NGC1068).  By comparison, the least mid-IR luminous maser is a kilomaser (M33).   
Thus, targeting more mid-IR luminous galaxies is bound to return a higher fraction of megamaser-emitting systems in general, and megamaser disks in particular.  

These differences may be suggestive of a stronger connection between a more significant dust heating in masers than in non-masers, however, that might be either proceeding more efficiently or may simply be quantitatively more enhanced in these sources.  Unfortunately, the average SED shapes do not offer a clear way of breaking this degeneracy; they simply reveal a clear difference between the masers and non-masers, with the latter having a significantly flatter SED in the longer mid-IR wavelengths, i.e., less re-processing by the dust; the slightly red average slopes we measure for the megamasers are not statistically significant different, which is consistent with the comparisons of average \emph{WISE} colors presented above. 
Nevertheless, the kilomasers are usually associated with star-forming regions (Lo 2005; Ott et al.\ 2013), and thus, for very similar mid-IR colors or SED slopes, the order of magnitude more power exhibited by megamasers is highly suggestive of an additional radiation source, e.g., an AGN.


Additionally, this comparison also suggests that $W1 - W4$ could also be exploited as a separator between masers and non-masers.   
Because the $W4$ band is the most luminous in megamasers, and especially for the disks, and is thus the wavelength region where the mid-IR emission peaks, the $L_{W4} = \nu L_{\nu}$ estimated at 22 $\micron$ can be an effective tool to explore for picking up the additional mid-infrared photons from the heated dust that is directly associated with the maser emission.   Indeed, when we compare the megamaser and disk detection rates in all four \emph{WISE} bands we find that they best correlate with $L_{W4}$.    We present in the following section a detailed analysis of the dependence of the maser properties on the $W1 - W4$ color, and show that it can be efficiently used as a measure of the obscured AGN activity in galaxy centers.



\subsection{Combining the Optical and mid-IR Information}

We are exploring here ways in which combinations of the optical and mid-IR photometric measurements and results presented in the previous sections can be best exploited towards identifying the closest links between maser emission and their host properties, and thus for enhancing the megamaser detection rates.

\subsubsection{ $W1-W4$ as a proxy for the $f_{\rm 24 \micron}/f_r$ flux ratio} \label{subsec:w14AGN}


As mentioned in Section~\ref{subsec:whywise}, the 24 $\micron$-to-optical ($f_{\rm 24 \micron}/f_r$) flux ratio of type 2 AGN is well correlated with the X ray--to--optical flux ratio, which makes the former parameter a good diagnostic for the ``relative" level of obscured AGN activity (Fiore et al. 2008; Georgantopoulos et al. 2008; Rovilos et al. 2011).  Because the optical flux of a galaxy can be converted to stellar mass ($M_{*}$) by assuming a universal initial mass function (Kauffmann et al. 2003a), and the X-ray flux can be used to estimate the total luminosity of an AGN ($L_{\rm AGN}$) by adopting a certain bolometric correction, the X ray-to-optical flux ratio is then a measure of the $L_{\rm AGN}/M_{*}$ ratio, or the specific black hole accretion rate $\lambda_{sBHAR}$.  This ratio also can also measure the rate of the black hole accretion normalized (or relative) to the host stellar mass (Aird et al. 2017; Georgakakis et al. 2017).   It has been argued that $\lambda_{sBHAR}$ is a more fundamental property of an AGN than the observed luminosity, and that, under certain assumptions, $\lambda_{sBHAR}$ can be a proxy for the Eddington ratio (e.g., Aird et al. 2017).

\begin{figure*}[ht] 
\begin{center} 
\vspace*{-1 cm} 
\hspace*{-1 cm} 
\includegraphics[angle=0, scale=0.47]{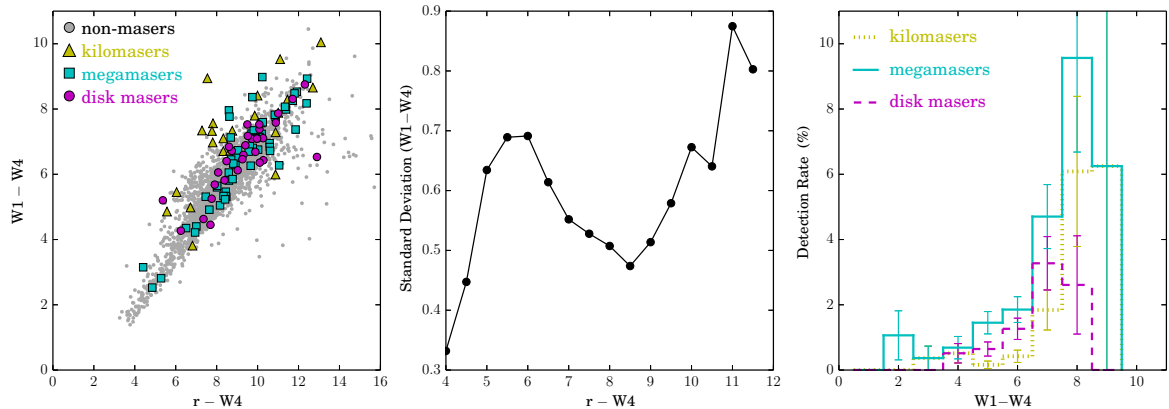}
\vspace*{0.0 cm} 
\caption{{\bf Left panel:} The distribution of the MCP galaxies in the ($r-W4$, $W1-W4$) parameter space.  The Spearman rank correlation coefficient between $r-W4$ and $W1-W4$ is 0.86. {\bf Middle panel:} The standard deviation of $W1-W4$ as a function of $r-W4$. {\bf Right panel:}  The kilomaser, megamaser and disk detection rates as a function of $W1-W4$.}
\label{w14}
\end{center} 
\end{figure*}

Because of the concurrence of the $W4$ band with the 24$\micron$ emission, the $r-W4$ color is a great proxy for the 24 $\micron$-to-optical flux ratio.  With the goal of exploring the greatly available  \emph{WISE} photometry for our MCP sources, and because the $W1$ band is the closest \WISE\ band to the optical $r$-band, we explore here the relationship between the $r-W4$ and $W1-W4$, and thus, the potential use of the $W1-W4$ as a measure of the ``relative" level of obscured AGN activity.  

We show in Figure ~\ref{w14} the distribution of all MCP galaxies with available SDSS $r$-band measurements in the $r-W4$, $W1-W4$ color-color diagram (left panel), along with the standard deviations of $W1-W4$ in each $r-W4$ bin (middle panel), and the detection rates of the various subtypes of masers as a function of $W1-W4$ (right panel).   It is quite apparent that $r-W4$ and $W1-W4$ correlate strongly (the Spearman's rank correlation coefficient is 0.86, and the standard deviations average at 0.6 and do not exceed 0.8) for the whole MCP sample. The correlation is only slightly weaker statistically when the kilo/mega/disk-maser sample are considered separately, the Spearman's rank correlation coefficient dropping to 0.6.  

The high degree of correlation of $r-W4$ with $W1-W4$ suggests that $W1-W4$ can act as an indicator of the level of the true AGN activity or the specific black hole accretion rate.   Under the assumption that the megamaser and the AGN activities are linked, this new tool should then offer significant improvements in the maser detection rates since only about half of our MCP sample has SDSS photometry while the $W1-W4$ color is measurable for nearly all of these objects.   Moreover, the maser detection rates, for all types of maser galaxies,  are clearly increasing with redder $W1-W4$ values, easily reaching greater than 5\% levels for $W1-W4 > 6$, which translates into a boost by a factor of $2-4$ relative to the current detection rates. 
Additionally, the nearby galaxies with $W1-W4 > 6$ account for 43\% of the total population, which corresponds to an improvement by a factor of three (or five) in the size of the potential parent target population when compared to the number of galaxies with $W1-W2 > 0.5$ (or $W1-W2 > 0.8$), regardless of the availability of SDSS photometry.   
Thus, $W1-W4$ has the potential to help enlarge substantially the sample size for studying the dependence of maser emission and its properties on the level or type of the associated AGN activity.


\subsubsection{$W1-W2$ versus $W1-W4$} \label{newWISEcolor}

\begin{figure}[ht] 
\begin{center} 
\vspace*{0 cm} 
\hspace*{0 cm} 
\includegraphics[angle=0, scale=0.47]{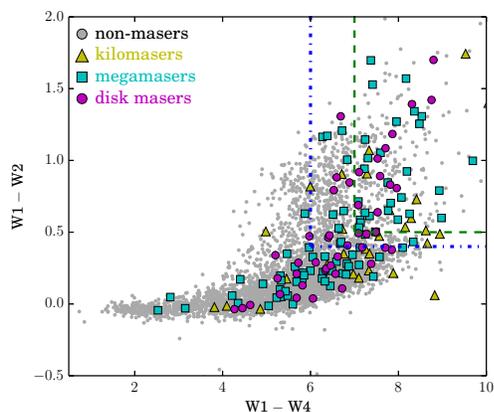}
\vspace*{0.0 cm} 
\caption{The distribution of the MCP galaxies in the ($W1-W4$, $W1-W2$) color-color diagram.  The green dashed lines correspond to $W1-W2 = 0.5$ and $W1-W4 = 7$. The overall maser detection rate for sample galaxies having $W1-W2 > 0.5$ and $W1-W4 > 7$ is 18$\pm$3\%, while megamasers and disks are expected to be detected at the impressive rates of $\sim 15\%$ and $\sim 5\%$ respectively.   A more relaxed constraint of $W1-W2 > 0.4$ and $W1-W4 > 6$ (shown with blue dot-dashed lines) still recovers high detection rates of 11\%, 9\%, and 3\% for the overall maser population, the megamasers, and the disks, respectively, with a 50\% increase in the completion rates.}
\label{newcolor}
\end{center} 
\end{figure}

Now, if $W1-W2> \sim0.5$ is effective for finding red/obscured AGN and a high fraction of masers, and since $W1-W4$ proves to be an adequate tool for picking up sources with a high ``relative" level of obscured AGN activity and also maser galaxies, it is of great interest to look into what these two colors can provide us with when working together.   We show in Figure ~\ref{newcolor} the distribution of the MCP galaxies in these two parameters. 
 
This plot primarily reveals that in the region where $W1-W2$ $>$ $\sim$0.5, adding the extra dimension of $W1-W4$ can effectively separate maser galaxies from non-maser galaxies. If we select a sample by using the criteria $W1-W2>0.5$ and $W1-W4 >7$,  the maser detection rate can be boosted to $18 \pm 2$\%, which is $\sim 6$ times higher than the current maser fraction (3\%).  Interestingly,  the overall maser detection rate does not show a further increase when  tighter selection criteria (e.g., W1$-$W2 $>$ 0.8) are considered.  Hence, 18\%  is about the maximum detection rate one can achieve, and that is obtained by using $W1-W4$ and $W1-W2$ to select maser emitting systems.    
In terms of of megamasers and their subgroup of disks, the detection rates associated with these criteria become 15$\pm$2\% and 5$\pm$1\%, respectively, which constitutes impressive $\sim 5$ times higher rates than the current values.

These rather dramatic increases in the maser detection rates are however shadowed by a relatively small pool of potential target galaxies, which translates into a high likelihood of missing a large fraction of megamasers and disks (only 32\% and 28\% of all megamasers and disks, respectively, fall under these criteria).    Maybe not  surprisingly, not all of the 18\% galaxies expected to exhibit red  \emph{WISE}  colors of $W1-W2>0.5$ show $W1-W4 >7$, leaving only 11\% of all galaxies in this parameter range.   Nevertheless, a slight relaxation of these constraints to $W1-W2 > 0.4$ and $W1-W4 > 6$ (shown in Figure ~\ref{newcolor} with blue dot-dashed lines) increases the parent target population to 17\%, and still recovers high detection rates of 11\%, 9\%, and 3\% for the overall maser population, the megamasers, and the disks, respectively.    The sample completion rates are also increased by 50\%, meaning that the likelihood of losing good megamasers (i.e., disks) is significantly diminished.   Thus, these criteria offer a very effective balance between high maser detection and completion rates, among selections based on optical classifications or mid-IR photometry, making this way of selecting maser targets the most promising for great enhancements in maser detections in future surveys.





\subsubsection{The combined effect of $L_{\rm MIR}$ and $W1-W4$} \label{subsec:mirlum_14}

\begin{figure*}[ht] 
\begin{center} 
\vspace*{0 cm} 
\hspace*{0 cm} 
\includegraphics[angle=0, scale=0.47]{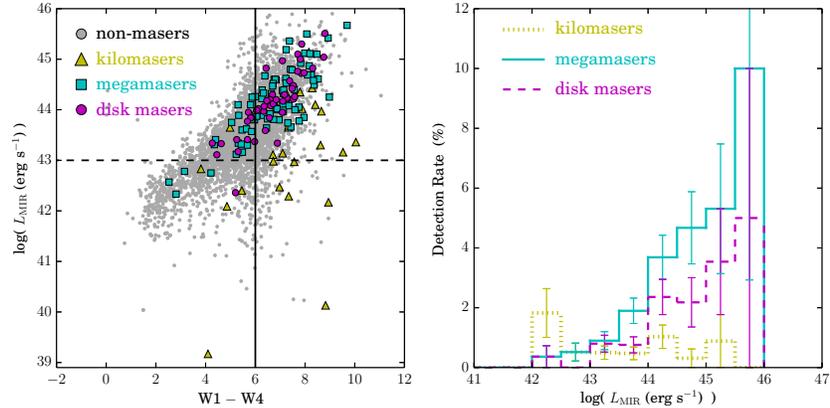}
\vspace*{0.0 cm} 
\caption{{\bf Left panel:} The distribution of the MCP galaxies in $L_{\rm MIR}$ and $W1-W4$. The Spearman-rank correlation coefficients are 0.67, 0.29, and 0.82, for the samples of all MCP objects, the kilomasers, and the megamasers, respectively.  {\bf Right panel:} The maser detection rates as a function of $L_{\rm MIR}$.}
\label{lum_14}
\end{center} 
\end{figure*}

Because in obscured AGNs a significant portion of the total AGN luminosity is re-emitted in the mid-IR wavelengths, if the strength of the AGN activity plays a role in exciting the maser emission, one might expect a correlation between the megamaser detection rate and their host's total mid-IR luminosity $L_{\rm MIR}$.  We explore this in Figure ~\ref{lum_14}, where we show the distribution of the MCP galaxies in the $L_{\rm MIR}$ and the $W1-W4$ color, along with the detection rates of various maser types as a function of $L_{\rm MIR}$.

Interestingly, while there is no obviously strong correlation between $L_{\rm MIR}$ and $W1-W4$ for the whole MCP sample or for the subsample of kilomasers,  when only the megamasers are considered the two mid-IR measures become rather tightly correlated: the Spearman rank coefficients are 0.67, 0.29, and 0.82, for the samples of all MCP objects, the kilomasers, and the megamasers, respectively.   Since $W1-W4$ can be an indicator of AGN strength (see Section ~\ref{subsec:w14AGN}), the correlation between $L_{\rm MIR}$ and $W1-W4$ for megamasers suggests that the mid-IR luminosity can also be employed to estimate the level of AGN activity, and thus, to enhance the maser detection rate.  Indeed, the megamaser and disk detection rates show a significant increase towards higher $L_{\rm MIR}$, with levels above 5\% for $L_{\rm MIR} \ga 10^{43}$ erg s$^{-1}$ where the kilomaser presence starts diminishing (Figure \ref{lum_14}, right panel).  

We note that this potential threshold in $L_{\rm MIR}$ for megamaser detection is reminiscent of the results of Zhu et al. (2011) who show that the maser detection rate increases as a function of the [\ion{O}{3}] luminosity, which is considered to be an important tracer of the bolometric luminosities of AGNs.  Thus, the rising maser fraction as a function of mid-infrared luminosity brings additional support to the scenario that the  AGN strength plays a significant role in exciting maser emission. 

In addition to strengthening the link between megamaser and AGN activities, this plot also indicates that the megamaser detection rate can be additionally boosted by at least a factor of $\sim$2 by selecting galaxies with $L_{\rm MIR}$ $\ge$ 10$^{44}$ ergs~s$^{-1}$, to $6.7 \pm 0.8$\% and $2.5 \pm 0.5$\% for megamasers and disks, respectively.   Furthermore, as we argue in the next subsections, by combining these trends with those in optical photometry we can improve the odds of finding megamaser disks even further.

\subsubsection{The combined effect of $L_{\rm MIR}$ and $u-r$} \label{subsec:mirlum_opt}


We recall that while the $f_{\rm 24 \micron}/f_r$  flux ratio in itself does not measure the AGN activity,  the ratio can be used to estimate the  level of AGN emission relative to that of the host (therefore, as discussed before, the relative level of AGN activity); thus, it is the enhancement in this ratio that reflects the strength of the putative AGN.   We also note that the $f_{\rm 24 \micron}/f_r$  ratios exhibit an intrinsic range among normal  galaxies (i.e., without a dominant AGN) that is highly dependent on the galactic morphology (e.g., Rovilos et al. 2011), with later types (of bluer optical colors) exhibiting higher intrinsic $f_{\rm 24 \micron}/f_r$ ratios than galaxies with earlier morphologies (of redder optical colors).   Thus, for a certain population of galaxies sharing the same (or no) level of AGN activity, the $f_{\rm 24 \micron}/f_r$ ratios increase as the galaxy colors change from red to blue without mirroring a significant (optically detected) AGN enhancement. 


\begin{figure*}[ht] 
\begin{center} 
\vspace*{0 cm} 
\hspace*{-1 cm} 
\includegraphics[angle=0, scale=0.47]{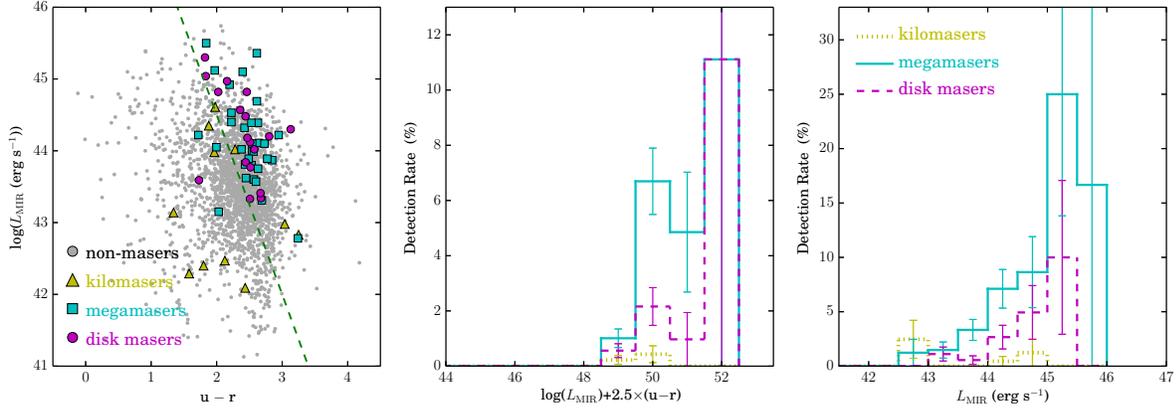}
\vspace*{0.0 cm} 
\caption{{\bf Left panel:} The distribution of the MCP galaxies as a function of $L_{\rm MIR}$ and $u-r$; only the galaxies with available SDSS photometry are plotted here.  The majority (88\%) of the megamasers lie above the green dashed line corresponding to log($L_{\rm MIR}) = -2.5 \times (u-r)+49.5$; {\bf Middle panel:} The maser detection rates as a function of log($L_{\rm MIR}) = -2.5 \times (u-r)+49.5$, for galaxies with log $(L_{\rm MIR}) > 43$, which includes all the currently detected disks with SDSS photometry; {\bf Right panel:} The maser detection rate as a function of $L_{\rm MIR}$ for sample galaxies having log($L_{\rm MIR}) = -2.5 \times( u-r)+49.5$.  }
\label{lum_opt}
\end{center} 
\end{figure*} 

We showed that the megamaser activity is more enhanced (i.e., higher megamaser detection rates) for stronger mid-IR emitters (higher $L_{\rm MIR}$ values), and with redder $W1-W4$ colors (Section \ref{subsec:mirlum_14}, Figure \ref{lum_14}), which also correlate well with higher $f_{\rm 24 \micron}/f_r$ ratios (Section \ref{subsec:w14AGN}, Figure \ref{w14}), while there is no significant correlation with the optical $u-r$ color, and thus with the host morphology (Section ~\ref{subsec:optical}, Figure \ref{optical}).   A natural conclusion of these trends is that the enhancement in the $f_{\rm 24 \micron}/f_r$ ratios in megamasers reflects AGN activity, and thus the megamaser excitation is linked to this type of nuclear activity.   
If megamasers are more easily excited when the AGN strengths increase beyond a certain threshold, one would expect the total mid-infrared luminosities of megamasers, which are correlated with the $f_{\rm 24 \micron}/f_r$ ratios, to increase as the optical galaxy colors become bluer.  We explore this possibility in Figure \ref{lum_opt}.

When the total mid-infrared luminosities of the MCP galaxies are plotted as a function of their $u-r$ colors (Figure \ref{lum_opt}, left panel), it becomes apparent that the megamaser number densities increase for larger $L_{\rm MIR}$ and redder $u-r$ values.   In fact, we find that most (90\%) of the megamasers lie above the line defined by $\log(L_{\rm MIR}) = -2.5 \times (u-r)+49.5$.    
Note that greater values of  $\log(L_{\rm MIR}) + 2.5 \times(u-r)$  essentially mean a greater degree of enhancement of the $f_{\rm 24 \micron}/f_r$  ratios by the AGN activity, and thus, once again, we are witnessing trends consistent with the idea that the AGN power may need to be above a certain high level to efficiently excite the megamasers.  

Moreover, the consideration of the optical $u-r$ color in conjunction with $L_{\rm MIR}$ appears to open new ways to improve on the maser detection rates.  When the maser detection rates are plotted as a function of $\log(L_{\rm MIR}) + 2.5 \times(u-r)$ (Figure \ref{lum_opt}, middle panel), keeping only  the galaxies with $\log (L_{\rm MIR}) > 43$, which includes all of the detected disks with optical (SDSS) photometry, we find that greater values of $\log(L_{\rm MIR}) + 2.5 \times (u-r)$ correspond to increasing enhancements in the megamaser and disk detection rates.    
In particular, for objects with log($L_{\rm MIR}) + 2.5 \times (u-r) > 49.5$ (that lie above the dashed green line in Figure ~\ref{lum_opt}, left panel) and $\log (L_{\rm MIR}) > 43$, the megamaser disk detection rates increase by at least a factor of 3 relative to that calculated for the entire sample of galaxies with available SDSS photometry,  mounting from 0.7\% to $\ga 2\%$, for 94\% completeness rates; for megamasers, the detection rate rises to 5.4\%, for a completion rate of 88\%.   For brighter mid-IR sources, with  $\log (L_{\rm MIR}) > 44$, the detection rates of all megamasers and disks in particular, increase to $\sim$8.3\% and 3.7\%, respectively, with corresponding completion rates of 54\% and 67\%.   These criteria manage to minimize the exclusion of known disk megamasers, while maximizing the inclusion of all types of megamasers.



%
\subsubsection{Enhancing the detection rate by including the $g$-band absolute magnitude } \label{subsec:gmag}


\begin{figure*}[ht] 
\begin{center} 
\vspace*{0 cm} 
\hspace*{-1 cm} 
\includegraphics[angle=0, scale=0.47]{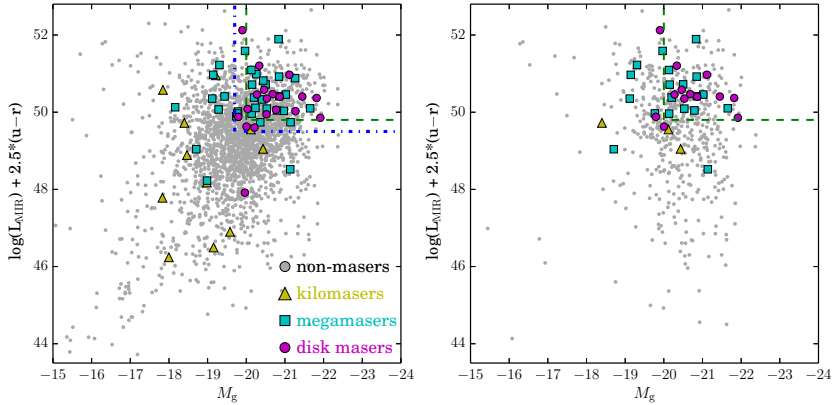}
\vspace*{0.0 cm} 
\caption{ {\bf Left panel:}  The distribution of the sample galaxies in the $M_{g}$ and log($L_{\rm MIR})+2.5\times (u-r)$ parameter space. The blue dashed lines correspond to $L_{\rm MIR}+2.5 \times(u-r) = 49.5$ and $M_{g} = -19.7$ while the green dashed lines indicate $L_{\rm MIR}$$+$2.5$\times$(u$-$r) $=$ 49.8 and $M_{g}$ $=$ $-$20.2;   
{\bf Right panel:} Same as the left panel, only for galaxies with $L_{\rm MIR} \ge 10^{44}$ ergs~s$^{-1}$.  }
\label{mir-gmag}
\end{center} 
\end{figure*} 

As seen in Section \ref{subsec:optical}, 100\% of the disk masers exhibit $g$-band absolute magnitudes $M_{g} < -19.7$, out of 69\% of all MCP galaxies with SDSS photometry, and thus it has potential for increasing both the target sample and the megamaser completion fraction.  Since $M_{g}$ is independent of $L_{\rm MIR}$ and $u-r$, including this parameter in the analysis adds a new independent dimension to studying the maser detection problem.

When plotting the $L_{\rm MIR}+2.5\times (u-r)$ parameter as a function of $M_{g}$ for all MCP galaxies with SDSS photometry (Figure \ref{mir-gmag}), it is readily apparent here that the majority of the megamasers  and disks lie in the top-right quadrant of the diagram.   In fact, by choosing $\log(L_{\rm MIR}) + 2.5 \times(u-r) \ge 49.5$ and $M_{g} < -19.7$ (marked in the figure by the dot-dashed blue lines), we select a sample of galaxies that includes 95\% of the disk megamasers and 79\% of all known megamasers.  The detection rates corresponding to these limits are 5.8$\pm$0.9\% and 2.5$\pm$0.6\% for the megamasers and disks, respectively, which is a  welcomed increase by $> 3$ times  relative to the current rates within the sample with SDSS photometric data  (i.e., 1.9$\pm$0.3\% for megamasers and 0.7$\pm$0.2\% for disk masers).


Interestingly, this particular enhancement in the maser detection can be improved even more by choosing just a slightly tighter cuts: for log($L_{\rm MIR}) + 2.5 \times(u-r) \ge 49.8$ and $M_{g} < -20.2$ (marked in the figure by the dashed green lines), we include 68\% of disk masers and 52\% of the known megamasers, while their detection rates skyrocket to 7.8$\pm$1.5\% and 3.7$\pm$1.0\%, respectively, which is an increase relative to the current rates by a factor of $\sim 5$.   
Thus, these criteria offer the highest megamaser and disk detection rates for the highest degree of completeness, especially for the disk systems, among all the criteria explored so far.   

Moreover, if we select a sample based on the method described above, but only among the galaxies with $L_{\rm MIR} \ge 10^{44}$ ergs~s$^{-1}$, the detection rates increase further to 11.9$\pm$2.5\% and 6.5$\pm$2.1\% (corresponding to an increase by a factor of $\sim 6 - 9$ in the detection rates) for megamasers and disks respectively, while the completeness rates reduce very little, to 44\% for megamasers and 58\% for the disks, which is still higher than when other criteria are employed (e.g., those targeting only \emph{WISE} red objects, with a conservative $W1-W2 > 0.5$ color cut, which recover only $< 30$\% of known megamasers or disks; see Section \ref{subsec:wisecolor}).



\subsection{The Dependence on the Isotropic Maser Luminosity}

The utility of megamasers in either understanding the physics of SMBH accretion or in making accurate distance determinations, and thus for cosmological constraints, is most efficient when these systems are bright enough so that a high enough sensitivity is achieved with a relatively short observing time.   It is therefore important to investigate the dependence of the total (isotropic) maser luminosities $L_{\rm H_2O}$ on the AGN/galactic properties.  

\begin{figure*}[ht] 
\begin{center} 
\hspace*{-1 cm} 
\includegraphics[angle=0, scale=0.47]{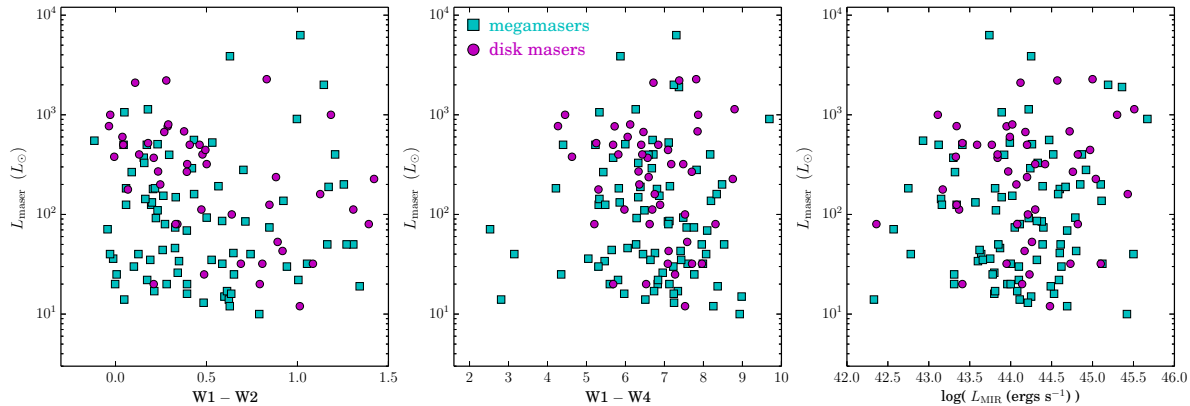}
\vspace*{0.0 cm} 
\caption{The isotropic maser luminosity $L_{\rm H_2O}$ as a function of $W1-W2$, $W1-W4$, and $\log (L_{\rm MIR}$). The filled cyan squares and the magenta circles represent megamasers and disk masers, respectively.}
\label{maserlum}
\end{center} 
\end{figure*}

Figure \ref{maserlum} illustrates the relation between $L_{\rm H_2O}$ and the  \emph{WISE} properties with the most potential association with strong maser emission, and thus with the highest megamaser detection rates, as discussed in the previous sections: the $W1-W2$ and $W1-W4$ colors (the left and middle panels, respectively), and the integrated mid-IR luminosity ($L_{\rm MIR}$).   At first glance, there are no apparent relationships here.   A closer look, however, reveals some interesting features.

A rather surprising trend is that the most luminous disk masers (i.e., those with $L_{\rm H_{2}O} > 300~ L_{\odot}$) appear to reside more preferentially in hosts with bluer \emph{WISE} colors, i.e., with $W1-W2 < 0.5$.  This could suggest that brighter disk masers prefer AGNs of bluer hosts, although this may simply be the result of the selection biases of the current maser samples: current maser surveys preferentially targeted optically known Seyferts (Constantin 2012) that are not optically obscured by high columns of obscuring dust, and thus, do not reprocess the central radiation at high enough levels that are seen as red in mid-IR.  The fact that, among all megamasers, there is no particular correlation of $L_{\rm H_{2}O}$ with $W1-W4$  or $L_{\rm MIR}$ supports the targeting bias scenario, and discards a possible correlation between the water luminosity $L_{\rm H_2O}$ and $W1-W2$. 
 
These trends do not necessarily contradict the previously suggested (albeit weak) dependence of the maser emission on the AGN strength by 
Zhu et al. (2011), Kondratko et al. (2006), and Castangia et al. (2013), who are employing the [\ion{O}{3}] and the X-ray luminosities as proxies for the AGN strength.
While our findings of higher megamaser detection rates for redder $W1-W4$ and stronger $L_{\rm MIR}$ suggest that an enhanced AGN activity eases the detection of a water maser system, an association between these two types of activity cannot be described by a simple correlation between their strengths. 

In addition to these mid-IR indicators of AGN activity, we also do not find any correlations between $L_{\rm H_2O}$ and any other combinations of optical and mid-IR properties we have explored so far.  This can be explained by recalling that megamasers often display significant variability and that the production rate of maser photons per unit volume depends on multiple physical parameters besides the strength of AGN, e.g., the amplification path lengths, the warping angle of an accretion disk (particularly for the disk masers; Neufeld, Maloney, \& Conger 1994), the temperature difference between gas and dust in the masing environment (e.g. Gary et al. 2016), the degree of saturation, etc., which can vary substantially in different systems.     We discuss in more detail these effects and their potential consequences on the correlations discussed so far in the next section.

\section{Discussion: The Relationship between the AGN Strength and the Megamaser Detection Rate}


This analysis provides new evidence for a strong relation between the megamaser detection rate $R_{\rm maser}$  
and a variety of optical and mid-IR photometric galactic properties that can act as proxies for the presence and likely the strength of the incumbent AGN.
Nevertheless, while we have uncovered ways of improving on the rate at which megamasers could be detected, from  a current $<1$\% to more than 15\%, the trends we revealed here do not provide us with new or satisfactory improvements on our understanding of the physics of water maser excitation processes.   

The correlations found to date between the detection rate as well as the strength and the morphology of the water maser emission and the nuclear properties of their host have been explored under the rather simple assumption that the megamaser emission  originates primarily from the thin accretion disk that is obliquely illuminated by a central X-ray source (e.g., Neufeld \& Maloney 1995, Neufeld 2000).   In this scenario, the total maser luminosity $L_{\rm H_{2}O}$ $\propto$ $L_{\rm AGN}$ (where $L_{\rm AGN}$ is the bolometric luminosity of the AGN), and thus a correlation between $R_{\rm maser}$ and $L_{\rm AGN}$ becomes reasonable because higher $L_{\rm AGN}$ leads to higher $L_{\rm H_2O}$, making megamasers easier to detect.

Indeed, we found that when compared to non-masers and kilomasers, the megamasers do appear to favor more luminous hosts, as well as stronger optical and mid-IR emitters.   However, among megamasers alone, there is no clear (or at most tentative) relation between the strength of the water maser emission and that of the   AGN activity, whether this is measured via optical photometry ($M_g$), optical line emission (e.g., $L_{\rm [O III]}$; Zhu et al. 2011; Constantin 2012), the X-ray activity (e.g., Castangia et al. 2013; Kondratko et al. 2006), or via its reprocessed counterpart in mid-IR by the surrounding  dust ($L_{\rm MIR}$ or $W1-W4$; this study).   

Particularly puzzling in this regard is the low luminosity nature of the AGN in NGC 4258. In this prototypical megamaser disk galaxy, the accretion rate of the supermassive black hole is estimated to be $10^{-4} M_{\odot}$~yr$^{-1}$ and the Eddington ratio is $\sim$1$\times$10$^{-4}$  (Herrnstein et al. 2005), showing that strong megamasers such as this one do not necessarily require high Eddington ratios. Therefore, it is not clear why $R_{\rm maser}$ diminishes significantly when the associated AGN becomes weaker.    Evidence for variability in both the AGN activity on multiple time scales (e.g., Ulrich et al. 1997) and that of the associated maser emission (e.g., Pesce et al. 2015) suggests that such variations, especially when they are not measured simultaneously in other wavelengths, may easily smear out a correlation between the AGN and megamaser strengths.    

It is pretty clear that more subtle physics must be involved, and we explore here two possible improvements in our understanding of these phenomena:  1) the potentiality of megamasers arising from or with the AGN outflows, and 2) the likelihood that megamasers originate from nearly edge-on sub-parsec warped gas disks.

\subsection{Connecting  Megamasers with AGN outflows}

AGN-driven gas outflows are generally good tracers of the AGN activity in general, and of the AGN feedback in action in particular (e.g., Rupke et al. 2005, Greene et al. 2012).   Powerful outflows are found to significantly influence the surrounding interstellar matter, and in particular, 
AGN outflows appear easily able to shock the circumnuclear gases leading to densities and temperatures that are suitably high for generating water maser emission.  For an efficient excitation of the 22 GHz water masers, the required number density and temperature of the molecular gas are $10^{8}-10^{10}$ cm$^{-3}$ and $400-1000$ K, respectively (Herrnstein et al. 2005; Gray et al. 2016).  

Maybe not unexpected is the fact that both the outflow rate of molecular gas and the power of AGN outflows are found to depend strongly on $L_{\rm AGN}$ (e.g., Fiore et al. 2017).   
 In the case of the ionized gas, the kinematic signature of AGN outflows is very common in luminous type 2 AGNs, showing
a clear increasing trend of the outflow fraction from 20\% to almost 100\% as the [\ion{O}{3}] luminosity increases from $10^{39}$ ergs s$^{-1}$ to $10^{42}$ ergs s$^{-1}$ (Woo et al. 2016; see also Kang et al. 2017 for H$\alpha$ line based AGN fractions).   Thus,  it is natural to speculate that as the AGN emission becomes stronger and the probability of finding powerful AGN outflow gets higher, the circumnuclear gas is more likely to get shocked and produce maser activity,  leading to the observed correlation between $R_{\rm maser}$ and AGN strength. 


While this simple picture is appealing, it remains practically unexplored.   Currently, high-resolution VLBI imaging of masers in jet systems only exists in a few galaxies, e.g., NGC 1052 (Claussen et al. 1998), Circinius (Greenhill et al. 2003), NGC 1068 (Gallimore et al. 2001), Mrk 348 (Peck et al. 2003), and thus, there are not enough statistical data to probe the above mentioned correlations.


\subsection{The Role of Maser Disk Warping}

That at least a fraction of megamasers originate in a disk-like configuration is now well established.    The real fraction of disk megamasers could easily be larger than what is currently observed as their detection requires that the gas disk be nearly edge-on (i.e., within a few degrees from being edge-on; Kuo et al. 2011; Reid et al. 2013; Kuo et al. 2013; Gao et al. 2016).   The nearly edge-on disk is necessary in order to provide the longest amplification path length that can lead to the strongest maser amplification.   
Previous maser surveys primarily targeted on Seyfert 2 galaxies mainly because of the higher likelihood of finding nearly edge-on disks in these systems, under the assumption that the standard model of AGN (Antonucci 1993) holds true.   
While the disk inclination is very likely the primary reason for the low maser disk detection rates, this parameter does not offer much interpretation for the correlation between $R_{\rm maser}$ and the strength of AGN, because it is mainly a geometrical effect.

To explain the dependence of $R_{\rm maser}$ on $L_{\rm AGN}$ for disk masers, one has to first understand how the necessary conditions for water maser excitation may depend on geometry itself.  
 Based on the standard theory of a thin steady-state accretion disk, the number density in the mid-plane of the disk can be written as a function of radial distance $r$ as (Herrnstein et al. 2005; Neufeld \& Maloney 1995):
 
\begin{equation}
n_{H_{2}} = {GM_{0} \dot{M} \over 3\pi(2\pi)^{1/2}\alpha r^{3}c_{s}^{3}m_{H_{2}}}~, 
\end{equation}

where $M_{0}$ is the central mass, $\dot{M}$ is the accretion rate, $\alpha$ is the Shakura-Sunyaev parameter of the kinematic viscosity (Shakura \& Sunyaev 1973), $c_{s}$ is the speed of sound, and $m_{H_{2}}$ is the mass of the hydrogen molecule. By assuming $L_{\rm AGN}$ $=$ 0.1$\dot{M}$c$^{2}$, where c is the speed of light, $\alpha=0.1$, and $c_{s}$ $=$ 2.3 km~s$^{-1}$ (corresponding to the gas temperature of 800 K), we can re-express $n_{H_{2}}$ as a function of the Eddington ratio $\eta_{\rm Edd}$ and $M_{0}$ as:

\begin{equation}
n_{H_{2}} = 7.8\times10^{8}\big(\frac{M_{0}}{10^{7}M_{\odot}}\big)\big(\frac{\eta_{Edd}}{10^{-4}}\big)\big(\frac{r}{0.1pc}\big)^{-3} ~~~cm^{-3}.
\end{equation}
From here, it is clear that as long as the gas can be heated to a few hundred kelvins, the density requirement for maser excitation can be easily fulfilled at highly sub-Eddington systems such as NGC 4258 at sub-parsec scales. 
By comparison with the needed gas density for maser excitation, what is actually more difficult to fulfill is the temperature requirement.  We also note, however, that, for NGC 4258, the assumption of steady-state accretion might not hold (Gammie, Narayan \& Blandford 1999).

In the standard picture of maser excitation in an accretion disk (Neufeld, Maloney, Conger 1994; Neufeld \& Maloney 1995), which is established based on the study of NGC 4258, the gas disk has to be warped so that the high-density mid-plane gas can be directly illuminated by the central X-ray source, to rise the gas temperature $T_{gas}$  above 400 K.  In a completely flat disk, the mid-plane is shielded from the heating source and the gas remains too cold (i.e. $T_{gas}$ $<$ 100 K) to mase; the viscous heating alone is insufficient to heat the disk to a sufficient temperature.  Therefore,  an external source of torque is needed in order to warp the disk so that the gas receives sufficient X-ray heating necessary to excite the maser.   In other words, \emph{the weaker the AGN is, the greater the degree of warping  needs to be in order for the gas to receive sufficient heating flux $F_{X}$}.

Disk warping plays yet another important role in producing or enhancing the maser emission.   After $T_{gas}$ rises above 400 K, the volume rate of maser photon production depends significantly on the temperature difference between gas and dust (Gray et al. 2016), which in turn depends primarily on $F_{X}$ (Neufeld, Maloney \& Conger 1994).  While these dependences are highly non-linear, mainly due to their dependence on other physical parameters such as pressure and gas density in the disk, there is tentative evidence for a correlation between the X-ray flux and the increase in the maser emissivity (Neufeld, Maloney \& Conger 1994; Neufeld 2000);  these calculations show that the amount of heating flux received by the gas in the disk increases with the obliquity of the disk (i.e., the tilt or the warping of the disk), and thus it is likely to result in higher maser emissivity.   A major caveat here is that the maser emissivity also depends on the bolometric luminosity of the central AGN:  while an increase in the X-ray luminosity can lead to an increase in maser emissivity, an increase in the bolometric luminosity that is not necessarily driven by an increase in X-ray luminosity will decrease the maser emissivity because the temperature difference between gas and dust becomes smaller.  Hence, the exact type of AGN activity will play a crucial role in determining the type of the emergent maser emission, and the exact inter-dependence remains a conundrum.

\subsection{The Probability of Finding Disks with Different Degrees of Warping}

Warping a gas disk within the gravitational sphere of influence ($R_{inf}$) of the central SMBH is difficult. There are only two mechanisms that have been demonstrated to be able to warp the sub-parsec maser disks effectively.  One way is via the Bardeen-Peterson effect (e.g. Caproni et al. 2007) caused by the interaction between the BH spin and the gas disk; however, this mechanism can only warp the disk in NGC 4258 by $\sim 3-4^{\circ}$ on a timescale of a few $\times$10$^{8}$ years (Bregman \& Alexander 2012) whereas the maser disk in this galaxy displays a warp angle of $\sim$8$^{\circ}$ (Herrnstein et al. 2005).  
A potentially much more efficient warping mechanism is the \emph{Resonant Relaxation} (Bregman \& Alexander 2009; Bregman \& Alexandar 2012), in which the gas disk is warped due to the residual torque produced by stars residing within $R_{inf}$.   In this situation,  the maser disk can be warped on a smaller timescale of $\sim 10^{7}$ years, however, the probability of producing warping angles of $\sim$10$^{\circ}$ is considerably smaller than that of warping angles of just a few degrees.

While the validity of the proposed warping mechanisms still needs to be constrained observationally, it is quite reasonable to conjecture that the probability of finding disks with smaller warping angles could be higher than that of finding disks with larger warping angles.  As a consequence, the maser detection rates $R_{\rm maser}$ will be expected to correlate with the AGN strength because as $L_{\rm AGN}$ increases the required degree of warping angles that would satisfy the masing conditions gets smaller, and hence, the probability of finding these disks becomes higher. 

We also note here that for luminous AGNs where the Eddington ratio $\eta_{Edd}$ gets above 0.1 another heating mechanism comes into play and the requirement of disk warping for maser excitation can become less stringent.   As demonstrated by the results of axisymmetric radiation hydrodynamic simulations of Namekata \& Umemura (2016), the mid-plane temperature of a completely flat dusty gas disk without warping around a BH of mass $10^7$ M$_{\odot}$ (i.e., the typical BH mass measured for maser disks) with $\eta_{Edd} = 0.77$, can rise above 400 K at the radius of 0.5 pc.  In this case, the heat transfer through collision between the gas in the mid-plane and the gas above/below the mid-plane which can be directly heated by the X-rays may become efficient enough so that the mid-plane gas can be warmed significantly without the disk being warped and illuminated directly.   Thus,  it is possible that the maser detection rate $R_{\rm maser}$ increases considerably for high $L_{\rm AGN}$ due solely to the fact that the temperature requirement for maser excitation becomes increasingly easier to fulfill.

To find evidence for these ideas, we compare the degree of warping $\theta_{\rm warp}$ (or how much the disk is bent) of the five disk maser systems (NGC 6323, NGC 6264, NGC 5765b, UGC 3789, and NGC 4258) that are found, to date, to have well defined disk structures in high sensitivity VLBI observations.  All of these systems have been modeled in three dimensions and the warping parameters are constrained with high accuracy (i.e., the position angles of the maser disk at different radii can be constrained to better than 2 degrees for systems such as NGC 4258).   We summarize the warping parameters and other basic parameters for each of these five galaxies in Table~\ref{tbl-warp}, where we sort them in increasing order of their $L_{\rm AGN}$.  Note that we do not include disk maser systems such as Circinus (Greenhill et al. 2003) in this analysis because these systems do not have acceleration measurements for their maser features which are required to constrain maser positions along the line of sight  in accurate three dimensional modelings for disk masers (Reid et al. 2013).

\begin{deluxetable*}{lrccrrl} 
\tabletypesize{\scriptsize} 
\tablewidth{0 pt} 
\tablecaption{Warping Angles  of Five Disk Maser Systems \label{tbl-warp}}
\tablehead{ 
\colhead{Galaxy} & \colhead{Distance } & \colhead{$L_{\rm AGN}$}  &  $\eta_{\rm Edd}$ & \colhead{$dp/dr$}  &  \colhead{$\theta_{\rm warp}$} & \colhead{Reference }  \\  
\colhead{Name}      & \colhead{(Mpc)}   & \colhead{(ergs~s$^{-1}$)} & \colhead{} & \colhead{(degree~mas$^{-1}$)}& \colhead{(degree)} & \colhead{}  
}     
\startdata 
NGC 4258  &    7.6 & $1.0\times10^{42}$  &  $2 \times10^{-4}$&    5.2$\pm$0.3 &     9.3$\pm$1.7  & Humphreys et al. (2013)    \\
NGC 6323  & 104.6 & $2.4\times10^{43}$ & 0.018 & 13.2$\pm$2.6 &  3.1$\pm$0.6    &   Kuo et al. (2015)          \\
UGC 3789  &  47.7  &  $2.5\times10^{44}$  &  0.13 &$-$2.0$\pm$1.1  &   1.0$\pm$0.6  &  Reid et al. (2013) \\ 
NGC 5765b& 120.7 & $3.3\times10^{44}$  & 0.02 & $-$3.5$\pm$0.4   &     0.7$\pm$0.1   &  Gao et al. (2016)     \\
NGC 6264  &  139.7 &  $5.6\times10^{44}$  &  0.132 &18.3$\pm$1.8  &   3.2$\pm$0.3  &  Kuo et al. (2013)  \\ 
\enddata 
\tablecomments{  
Column 1: Name of the galaxy; Column 2: The distance to the galaxy used for converting the angular scale to the physical scale; Column 3: The bolometric luminosity of the AGN.  With the exception of NGC 5765b, the bolometric luminosities are from Greene et al. (2010).  For NGC 5765b, the bolometric luminosity is estimated from the [\ion{O}{3}] luminosity reported in Gao et al. (2017), with the bolometric correction from Liu et al. (2009); Column 4:  The Eddington ratios; Column 5: The change of the position angle with the angular radius. These are the best-fit warp parameters from three-dimensional Bayesian modeling of the observed maser disks, with the corresponding references listed in Column 7;  Column 6: The warping parameter of the maser disk, $\theta_{\rm warp}$, which is evaluated by  $\theta_{\rm warp}
=dp/dr(r_{in}-r_{out})+d^{2}p/dr^{2}(r_{in}^{2}-r_{out}^{2})$, where $r_{in}$ and  $r_{out}$ are 0.17 and 0.37 pc, respectively;  $d^{2}p/dr^{2} = -0.24 \pm 0.02$ for NGC 4258, and it is unavailable for the other galaxies except for UGC 3789 where this is set to zero given that the warp is extremely  small.   Column 7: The references from which we obtained the best-fit warp parameters.  
}
\end{deluxetable*}

The degree of disk warping, $\theta_{\rm warp}$, is generally defined as the change in the position angle of the disk from the inner radius $r_{\rm in}$ to the outer radius $r_{\rm out}$ of the disk.  We note however that different maser disks usually exhibit different  $r_{\rm in}$ and $r_{\rm out}$, which makes any comparison of the warping quite challenging.     For the sake of consistency, we re-define the $\theta_{\rm warp}$ as the change in the position angle of the disk over a radial extent of 0.2 pc, stretching from  $r_{\rm in} = 0.17$ pc to $r_{\rm out} = 0.37$ pc;  $r_{\rm in}$ $=$ 0.17 pc is the inner radius of the maser disk in NGC 4258 (Herrnstein et al. 2005).

Among these five galaxies, NGC 4258 has the lowest bolometric luminosity and Eddington ratio ($L_{\rm AGN} = 10^{42}$ ergs~s$^{-1}$; $\eta_{\rm Edd} = 2\times10^{-4}$), and the largest $\theta_{\rm warp}$ $=$ 9.3$^{\circ}\pm1.7^{\circ}$ (Humphreys et al. 2013).   As $L_{\rm AGN}$ increases by two orders of magnitude, $\theta_{\rm warp}$ becomes significantly smaller, by at least a factor of three and reaching even an order of magnitude.   Unfortunately, the Eddington ratios of the four more luminous maser disks are similar, and thus are not particularly helpful in identifying additional trends with the degree of warping.   

From these comparisons, one can easily see that as the AGN strength increases, $\theta_{\rm warp}$ drops and can become quite negligible at the highest AGN luminosities among our sample. In light of a possible strong link between the AGN and the water masing strengths, this would translate into less warped maser disks found with maser searches targeting objects with higher Eddington ratios.  
While the number statistics are currently too small to confirm any clear relation between the maser disk warping and the AGN strength,  as well as provide sensible constraints of the detection rates of maser disks of a certain degree of warping, the general trend is well consistent with the proposed picture.    If and when more high quality images of disk maser systems become available, we will be in a better position to test and quantify this proposed relationship between the warping angles of the disk masers and the AGN strength and the accretion efficiency, and thus should provide crucial insights into the link between the megamaser excitation and SMBH accretion disks.\\



\section{Conclusions: Guidance for Future Maser Surveys }

To reach the ultimate potential of H$_{2}$O megamasers for measuring a percent-level \ho, as well as for a significant increase in the number of accurately measured  masses of SMBHs, we need efficient searches for new megamaser disk systems.   We address here ways in which we could greatly improve our quest for finding these exotic systems, by exploring the maser host galaxy characteristics that relate to, and thus possibly nurture the production of powerful  masers in their centers.  

Using data from the \emph{Wide-Field Infrared Survey Explorer} (\emph{WISE}), with optical photometry from SDSS as auxiliary information, we present here a thorough analysis of the photometric properties of the largest sample of galaxies surveyed for H$_2$O maser emission in 22 GHz, with the goal of better constraining the previously proposed connection between water masing activity and the circumnuclear dust absorption and radiation reprocessing in galaxy centers.   To date, this kind of information is scarce, as no systematic study of the multi-wavelength properties of the galaxies observed in \h2o ~ maser emission at 22 GHz has been performed for a sound statistical sample.  We only now have the statistics to pursue such an investigation.   

Our main results are summarized in Table ~\ref{tbl-summary}, and are briefly discussed as follows:

\begin{deluxetable*}{lcccccccc} 
\tablewidth{0 pt} 
\tablecaption{Effective Search Criteria for Megamasers \& Disks \label{tbl-summary}}
\tablehead{ 
\colhead{Search} & \colhead{\% All MCP} & \colhead{$R_{\rm maser}$} & \colhead{C$_{\rm maser}$} & \colhead{$R_{\rm Mmaser}$} & \colhead{C$_{\rm Mmaser}$} & \colhead{$R_{\rm disk}$} & \colhead{C$_{\rm disk}$} & \colhead{$R^{100-300}_{\rm disk}$($N_{\rm disk}$)\tablenotemark{a} }\\
\colhead{Criteria}      & \colhead{Galaxies}   & \colhead{\%} & \colhead{\%} & \colhead{\%} & \colhead{\%} & \colhead{\%} & \colhead{\%} & \colhead{ }
}     
\startdata 
&  &  &  &  &  &  &  & \\
WISE (SNR $>$ 3) &  86  & 3.3$\pm$0.3 & 100 & 2.7$\pm$0.2 &  98&  0.9$\pm$0.1 & 100 & \nodata \\
&  &  &  &  &  &  & &  \\
\hline
&  &  &  &  &  &  & &  \\
SDSS photometry  &   57  & 2.3$\pm$0.3 & 40 & 1.9$\pm$0.3 &  40&  0.7$\pm$0.2 & 40& 0.7\% (163)\\
&  &  &  &  &  &  &  & \\
\hline
&  &  &  &  &  &  &  & \\
$W1-W2 > 0.5$ &  18 & 8.8$\pm$1.1 & 45 & 7.0$\pm$1.0 &  44 &  2.3$\pm$0.5 & 39& 0.9\% (3)\\
&  &  &  &  &  &  & &  \\
\hline
&  &  &  &  &  &  & &  \\
$W1-W2 > 0.8$ &  9&  9.5$\pm$1.6 & 25 & {\bf 8.0$\pm$1.4} &  26 &  {\bf 3.3$\pm$0.9} & 30& 1.8\% (1)\\
&  &  &  &  &  &  & &  \\
\hline
$W1-W2 > 0.5$ &  &  &  &  &  &  &  & \\
$W1-W4 > 7$ &  6 &  18.2$\pm$2.5 & 32 & {\bf14.9$\pm$2.3} &  32 &  {\bf 4.7$\pm$1.3} & 28& 2.8\% (3)\\
&  &  &  &  &  &  &  \\
\hline
$W1-W2 > 0.4$ &  &  &  &  &  &  &  & \\
$W1-W4 > 6$ & 17 &  10.5$\pm$1.2 & 48 & {\bf 8.5$\pm$1.1} &  {\bf 48} &  {\bf 3.1$\pm$0.6} & {\bf 49}& 1.4\% (5)\\
&  &  &  &  &  &  & &  \\
\hline
&  &  &  &  &  &  & &  \\
$L_{\rm MIR} > 10^{44}$ ergs~s$^{-1}$  & 27 &  7.4$\pm$0.8 & 53 & {\bf 6.7$\pm$0.8} &  {\bf 58} &  {\bf 2.5$\pm$0.5} & {\bf 62}& 1.6\% (36)\\
&  &  &  &  &  &  &  & \\
$L_{\rm MIR} > 10^{43}$ ergs~s$^{-1}$  & 76 &  4.4$\pm$0.4 & 90 & 3.8$\pm$0.3 &  {\bf 96} &  1.4$\pm$0.2 & {\bf 98}& 0.9\% (170)\\
&  &  &  &  &  &  &  & \\
\hline
&  &  &  &  &  &  &  & \\
$\log(L_{\rm MIR}) + 2.5(u-r) > 49.5$ &  &  &  &  &   &   & &\\
$L_{\rm MIR} > 10^{43}$ ergs~s$^{-1}$  & 44 &  4.8$\pm$0.7 &  79  &  4.4$\pm$0.7 &   90  &  1.7$\pm$0.2  & 94 & 1.5\% (67)\\
&  &  &  &  &  &  &  & \\
\hline
&  &  &  &  &  &  &  & \\
$\log(L_{\rm MIR}) + 2.5(u-r) > 49.8$ &  &  &  &  &  &  &  & \\
$M_{\rm g} < -20.2$  & 14 &  7.8$\pm$1.5 & 43 & {\bf 7.8$\pm$1.5} &  {\bf 52} &  {\bf 3.7$\pm$1.0} & {\bf 68} & 3.3\% (81)\\
&  &  &  &  &  &  &  & \\
\hline
$\log(L_{\rm MIR}) + 2.5(u-r) > 49.8$ &  &  &  &  &  &  &  & \\
$M_{\rm g} < -20.2$  &  &  &  &  &  &  & & \\
$L_{\rm MIR} > 10^{44}$ ergs~s$^{-1}$  &  6 & 11.8$\pm$2.8 & 29 & {\bf 11.8$\pm$2.8} &  {\bf 44} &  {\bf 6.5$\pm$2.1} & {\bf 58}  & 5.6\% (20)\\
\enddata 
\tablecomments{ $R_{\rm maser}$, $R_{\rm Mmaser}$, and $R_{\rm disk}$ are the detection rates of all masers, megamasers, and disks, respectively.   C is the completeness rate, listed separately for all masers, megamasers, and disks.  The highest detection and completion rates for megamasers and disks are highlighted in bold.  Note that the detection rates and completeness rates shown in the last three rows of this table are calculated relative to the entire sample of galaxies with available SDSS photometry. } 
\tablenotetext{a}{$N_{\rm disk}$ = Predicted number of new disk masers that could be detected based on each criteria, from the SDSS-DR14 sample of nearby galaxies with photo-$z < 0.1$, photo-$z$Err $ < 0.01$ , photo-$z$Err/photo-$z < 1/3$  and distance between 100 and 300 Mpc (a total of 23277 galaxies); $R^{100-300}_{\rm disk}$ = The disk detection rate prorated for this particular distance range.}
\end{deluxetable*}

\begin{itemize}

\item {\tt  Red $W1-W2$ colors:} The mid-IR color cuts for selecting red/dusty AGNs reveal a dramatic increase in the maser detection rates (up to $\sim 10\%$ for all masers, and $\sim 7\%$ for the megamasers and disks, depending on the exact criteria involved in defining the red \emph{WISE} galaxies, which is a boost by at least a factor of four relative to the current rates).   
Nevertheless, these criteria, which link once again the masing activity with dusty/obscured black hole accretion,  correspond to a rather small maser completeness rate: they would miss the bulk of the megamaser disks out there, as the majority ($\sim$60\%) of the masers and disks are found to be hosted by \emph{WISE} blue galaxies.    Also, the potential pool of such galaxies remains small, as only 16\% of our entire sample satisfy this criterion.

\item {\tt Red $W1-W2 $ \& $W1-W4$ colors}:  By adding the $W1-W4$ information we are able to avoid galaxies with low emission in $W4$ at 22$\micron$ where the maser detection rates are the lowest, and to raise the detection rates of all masers and megamasers to $\sim 18$\% and $\sim 15$\%, respectively, depending on how stringently red are considered the galaxies.  The completeness rate can be higher than 48\% for detection rates that are three times better than the current ones. 
While with the most stringent red colors the detection rates can be the highest among all other sample selection methods we have explored here, they would be able to target only 30\% of the megamasers, meaning that we would still miss the bulk of the maser disks.  
Because there are only $\sim 6$\% of galaxies that comply with these criteria, these surveys are less prohibitive in terms of the total observing time.  

\item {\tt $L_{\rm MIR} > 10^{44}$ ergs~s$^{-1}$}:   By selecting galaxies with luminous mid-IR emission, the maser detection rates can be increased by a factor of $2-3$ from the current ones.  While the maser detection rates are lower in comparison with the previous two methods, they are still a factor of at least three larger than the current  ones, with a rate of completeness that reaches $\sim 60\%$ for megamasers and disks.   For a slightly lower  mid-IR luminosity, i.e., $L_{\rm MIR} > 10^{43}$ ergs~s$^{-1}$, the pool of available galaxies to be surveyed increases by a factor of three, and the disk completion rates rises to $\sim 100\%$ for megamaser and disk detection rates that almost double the current detections.  


\item {\tt $\log(L_{\rm MIR}) + 2.5\times(u-r) > 49.5$ and $L_{\rm MIR} > 10^{43}$ ergs~s$^{-1}$}:  A combination of intrinsically bright mid-IR emission and optical colors results in reasonably high detection rates (double the current ones) with only a small likelihood of not surveying disks or megamasers in general (the completion rates are $> 90\%$).  These criteria are particularly useful in avoiding the kilomasers, targeting most efficiently the megamaser activity, with or without an obvious disk.   

\item {\tt $\log(L_{\rm MIR}) + 2.5\times(u-r) > 49.8$ and $M_{\rm g} < -20.2$}:  By combining the total mid-IR luminosity with optical colors and optical brightness, the maser detection rates rise to $\sim 8$\% and $\sim 4$\% for the megamasers and disks, respectively; this represents an enhancement of a factor of 4--5 times relative to the current ones, among all surveyed galaxies with SDSS photometry.   
While the detection rate of megamasers is comparable to those drawn from samples of galaxies with $W1-W2 > 0.8$, the completeness is  increased 3-fold,  especially for the disk masers: 52\% of megamasers and 68\% of disk masers satisfy these criteria.   The total number of galaxies in this parameter range is also two times larger than in the case of $W1-W2 > 0.8$ limit, increasing thus the total number of potential disk detections.

\item {\tt $\log(L_{\rm MIR}) + 2.5\times(u-r) > 49.8$ and $M_{\rm g} < -20.2$ plus $L_{\rm MIR} > 10^{44}$ ergs~s$^{-1}$}:   When an additional condition on the integrated mid-IR brightness is employed, the megamaser and disk detection rates increase to $\sim 12$\% and $\sim 6$\%, respectively (a factor of $6-9$ increase relative to the current rates), while the completeness remains relatively high (i.e., above 44\% or higher).  These criteria should constitute the most efficient way of hunting for the next megamaser disks.   

\end{itemize}

Of course, one of the most pressing questions is finding how many un-surveyed galaxies are in these particular parameter spaces, and how many megamasers might they detect.  What really matters for improving \ho ~ or SMBH statistics is not so much the completeness as the sheer numbers.   
While previous maser surveys might have somewhat exhausted the main samples of nearby spectroscopically identified galaxies, there is a significantly higher number of galaxies with photometric redshifts that have not yet been explored for water maser emission.    

Working with, for example, the SDSS DR14 sample of nearby galaxies with photometric redshifts, a best match to the redshift (or distance) distribution of the MCP samples discussed here is given by the selection of all galaxies with photo-$z < 0.1$, photo-$z$Err $ < 0.01$ , photo-$z$Err/photo-$z < 1/3$  and distance between 100 and 300 Mpc (for objects closer than 100 Mpc the photometric redshifts are not reliable).

Tailoring our analysis for this particular distance range (which results in slightly smaller detection rates for maser galaxies of all types), and applying the photometric criteria summarized above, we find that quite a high number of megamaser disk systems are still to be detected.    These number predictions are listed in the last column of Table~\ref{tbl-summary}, which reflects the potential to find 20 (or 81) new maser disks from a survey of carefully chosen  $\sim 360$ (or $\sim 2500$) galaxies, that obey the last (or the second to last) set of criteria listed in Table~\ref{tbl-summary}, making these results supremely valuable.    We caution, however, that this success rate may be subject to a preselection on strong narrow line emission activity, as it was generally employed in building the MCP samples.

Regarding the need and the use of the new hopeful detections of megamaser disks, a 1\% accuracy in \ho\  can potentially be achieved via either: i) the measurement of $\sim$10\% accuracy distances to $ \ga100$ disk maser systems, or  ii) the measurement of high-accuracy (~3\%) distances to $\sim$10 disk maser galaxies (e.g., Greenhill et al. 2009).    The number of disk masers from which one can obtain distances with $\sim$10\% accuracy is likely to remain below 100 in the following decade, even when applying our most efficient predictions for selection (Table ~\ref{tbl-summary}), as the survey time would become prohibitively long, making the first option utterly impractical.  While the current \emph{VLBI} sensitivity can provide the needed 3\% accuracy only for one maser disk (NGC 4258), the Next Generation Very Large Array (\emph{ngVLA}) within the \emph{VLBI} is bound to make that accuracy possible for a larger number of maser disks, making the second option for achieving 1\% \ho\ actually feasible:  
the current maser surveys provide a fraction of $\sim 1/7$ disk masers with sufficient quality for 3\% accuracy in the  distance measurements, suggesting that all we need is the discovery of $\sim$70 new maser disks; with the $\sim$6\% detection rate we predict here, we will need to survey $\sim$1200 narrow-emission line galaxies that satisfy the criterion, and this is doable with a \emph{GBT} scanning time of $\sim$10 minutes per galaxy, or a total of 200 \emph{GBT} hours (which is the typical \emph{GBT} time awarded to the MCP per year before 2013).



The work presented here constitutes our first step to enhance the maser detection rate.   We find that, although employing mid-IR criteria improves considerably the megamaser detection rates, the mid-IR emission alone does not seem to be particularly sensitive or constraining to the maser pumping mechanism.  On the other hand, the maser disk detection could remain the only way to acknowledge accreting SMBHs in \emph{WISE} blue hosts whose nuclei do not show clear signs of AGN-like optical nebular properties, and thus could provide an improved way to compute a refined census of active black holes in galaxy centers.   
A better understanding of the individual contributions from the host starlight, the nuclear star formation, as well as the central AGN to the mid-IR properties, and thus to the effects of each of these activities on the water masing processes, can be achieved with SED fittings of the light emitted by the hosts of the various subtypes of masers and of non-masers, and we are presenting this analysis in a future paper (Kuo et al. 2019, in preparation).




\acknowledgements

We gratefully acknowledge the anonymous referee for a very thorough and insightful review that improved this manuscript.    
This publication is supported by Ministry of Science and Technology, R.O.C. 
under the project 104-2112-M-110-014-MY3.   
A.C. acknowledges support from 4-VA, a collaborative partnership for advancing the Commonwealth of Virginia, along with the hospitality of the The National Radio Astronomical Observatory's North American ALMA Science Center in Charlottesville, VA, where part of this work was done.  
J.-H.W. acknowledges support by the Basic Science Research Program through the National Research Foundation of Korea government (No. 2017R1A5A1070354).   

This work has made use of data products from the \emph {Wide-field Infrared Survey Explorer (WISE)} and the SDSS.
\emph {WISE} is a joint project of the University of California, Los Angeles, and the Jet Propulsion Laboratory/California Institute of Technology, funded by the National Aeronautics and Space Administration.   
SDSS is managed by the Astrophysical Research Consortium for the Participating Institutions of the SDSS-III Collaboration including the University of Arizona, the Brazilian Participation Group, Brookhaven National Laboratory, Carnegie Mellon University, University of Florida, the French Participation Group, the German Participation Group, Harvard University, the Instituto de Astrofisica de Canarias, the Michigan State/Notre Dame/JINA Participation Group, Johns Hopkins University, Lawrence Berkeley National Laboratory, Max Planck Institute for Astrophysics, Max Planck Institute for Extraterrestrial Physics, New Mexico State University, New York University, Ohio State University, Pennsylvania State University, University of Portsmouth, Princeton University, the Spanish Participation Group, University of Tokyo, University of Utah, Vanderbilt University, University of Virginia, University of Washington, and Yale University. This research has made use of the NASA/IPAC Extragalactic Database (NED) which is operated by the Jet Propulsion Laboratory, California Institute of Technology, under contract with the National Aeronautics and Space Administration. Funding for SDSS has been provided by the Alfred P. Sloan Foundation, the Participating Institutions, the National Science Foundation, and the U.S. Department of Energy Office of Science. The SDSS web site is http://www.sdss.org/.

\end{document}